\documentclass[aps,prx,superscriptaddress,reprint]{revtex4-1}
\usepackage{amsmath}
\usepackage{amssymb}
\usepackage{graphicx}
\usepackage{url}
\usepackage{hyperref}
\usepackage{color,xcolor}
\usepackage{epstopdf}
\usepackage{float}
\usepackage{ulem}
\usepackage[percent]{overpic}

\begin{document}

\title{Oxygen magnetic polarization, nodes in spin density, and zigzag spin order in oxides}

\author{Ling-Fang Lin}
\affiliation{Department of Physics and Astronomy, University of Tennessee, Knoxville, Tennessee 37996, USA}
\author{Nitin Kaushal}
\affiliation{Department of Physics and Astronomy, University of Tennessee, Knoxville, Tennessee 37996, USA}
\affiliation{Materials Science and Technology Division, Oak Ridge National Laboratory, Oak Ridge, Tennessee 37831, USA}
\author{Cengiz \c{S}en}
\affiliation{Department of Physics, Lamar University, Beaumont, Texas 77710, USA}
\author{Andrew D. Christianson}
\affiliation{Materials Science and Technology Division, Oak Ridge National Laboratory, Oak Ridge, Tennessee 37831, USA}
\author{Adriana Moreo}
\author{Elbio Dagotto}
\affiliation{Department of Physics and Astronomy, University of Tennessee, Knoxville, Tennessee 37996, USA}
\affiliation{Materials Science and Technology Division, Oak Ridge National Laboratory, Oak Ridge, Tennessee 37831, USA}

\begin{abstract}
Recent studies on Ba$_2$CoO$_4$ (BCO) and SrRuO$_3$ (SRO) have unveiled a variety of intriguing phenomena, such as magnetic polarization on oxygens, unexpected nodes in the spin density profile along bonds, and unusual zigzag spin patterns in triangular lattices. Here, using simple model calculations supplemented by DFT we explain the presence of nodes based on the antibonding character of the dominant singly occupied molecular orbitals along the transition metal (TM) to oxygen bonds. Our simple model also allows us to explain the net polarization on oxygen as originated from the hybridization between atoms and mobility of the electrons with spins {\it opposite} to those of the closest TM atoms. Our results are not limited to BCO and SRO but they are generic and qualitatively predict the net polarization expected on any ligands, according to the spin order of the closest TM atoms and the number of intermediate ligand atoms. Finally, we propose that a robust easy-axis anisotropy would suppress the competing 120$^{\circ}$ degree antiferromagnetic order to stabilize the zigzag pattern order as ground state in a triangular lattice. Our generic predictions should be applicable to any other compound with characteristics similar to those of BCO and SRO.
\end{abstract}

\maketitle

\section{Introduction}
Due to the interplay of charge, spin, orbital, and lattice degrees of freedom, the exotic electronic and magnetic properties of strongly interacting electrons, particularly transition metal (TM) oxides, have attracted broad interest over decades in the Condensed Matter community~\cite{scalapino2012common,fradkin2015colloquium,Dagotto:Prp,Dagotto:Rmp94}. Remarkable phenomena have been unveiled, such as high critical temperature superconductivity in copper-, iron-, and nickel-based materials~\cite{bednorz1986possible,Dagotto:Rmp94,Kamihara:Jacs,Dai:Np,li2019superconductivity,zhang2020similarities}, colossal magnetoresistance and phase separation in manganites~\cite{Dagotto:Prp,Salamon:Pmp,kusters1989magnetoresistance,zhu2020nonmonotonic,miao2020direct}, orbital ordering in perovskites~\cite{tokura2000orbital,varignon2019origins,pandey2021origin}, spin block states~\cite{herbrych2019novel,herbrych2020block,herbrych2020block1,zhang2020iron}, orbital-selective Mott phases~\cite{patel2019fingerprints}, ferroelectricity~\cite{Scott:Science,Cheong:Nm,lin2019frustrated,zhang2021peierls} and several others.

In the many oxide materials, the anions O$^{2-}$ play important roles, not only as ligands that bind to the central metal atoms when they form coordinated polyhedrons, but also as bridges that connect these polyhedrons by either corner-, edge-, or face-sharing geometries. In principle, these closed-shell anions should not be magnetic, such as in the well-known superexchange and double exchange models~\cite{anderson1950antiferromagnetism,Anderson:PR}. In several simplified calculations, their only influence resides in the actual value of the electronic hopping amplitudes in eV units, such as when one-orbital Hubbard models are used for cuprates.

However, recent polarized neutron diffraction experiments unveiled the surprise that a robust magnetic polarization is present in all oxygen sites -- contributing 30\% of the total magnetization -- in the case of the metallic ferromagnet SrRuO$_3$~\cite{kunkemoller2019magnetization}. Previous studies also found a nonzero magnetization density at the oxygen ions in the yttrium iron garnet~\cite{bonnet1979polarized}, Li$_2$CuO$_2$~\cite{chung2003oxygen}, La$_{0.8}$Sr$_{0.2}$MnO$_3$~\cite{pierre1998polarized}, YTiO$_3$~\cite{kibalin2017spin}, and Ca$_{1.5}$Sr$_{0.5}$RuO$_4$ \cite{gukasov2002anomalous}. All these materials contain transition metal spins coupled into a global ferromagnetic (FM) state. A small magnetization density at the oxygens can also be visually inferred from the polarization neutron diffraction figures reported for Sr$_2$IrO$_4$ in a canted antiferromagnetic (AFM) state~\cite{jeong2020magnetization}.

\begin{figure}
\centering
\includegraphics[width=0.42\textwidth]{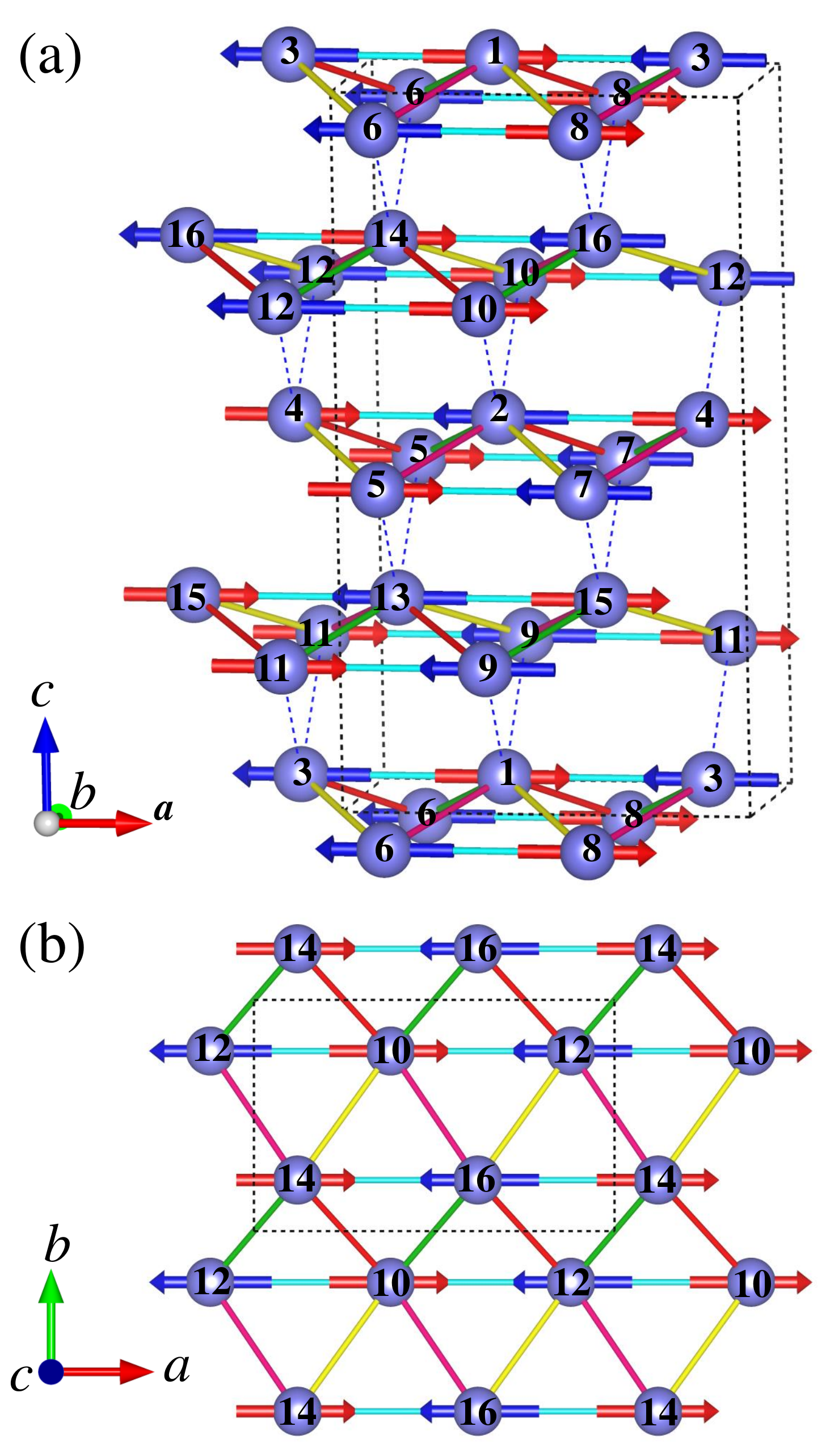}
\caption{(a) Three-dimensional illustration of the Co-Co bonds in BCO and the simplified AFM collinear structure (dubbed AFM0). (b) Top view of the AFM0 state in the $ab$-plane projection. The black dashed lines define the magnetic unit cell. Red and blue arrows represent opposite directions of the Co magnetic moments. The numbers label different locations of Co atoms in the magnetic unit cell, while Ba and O atoms are not shown here for simplicity. The inequivalent paths between adjacent Co atoms are also displayed with bars in different colors (blue, red, green, yellow, and magenta). Note that the real magnetic structure is slightly noncollinear, while the AFM0 state constructed here is collinear for simplicity.}
\label{mag}
\end{figure}

In addition, as a result of the covalent hybridization between TM and O atoms, the reduction of the magnetic moment on the TM atoms was reported in AFM materials, such as Sr$_2$CuO$_3$~\cite{walters2009effect}, Ca$_3$Fe$_2$Ge$_3$O$_{12}$ garnet~\cite{plakhty1999spin}, and Cr$_2$O$_3$~\cite{brown2002determination}. All these findings challenge our conventional assumption regarding the passive role of oxygen, and a revaluation of the importance of the ligand for magnetic exchange interactions in these TM oxides is required.

Ba$_2$CoO$_4$ (BCO) provides a more recent example highlighting again the unusual role of oxygen. Different from other well-explored cobalt oxide compounds, BCO, isostructural to $\beta$-Ca$_2$SiO$_4$, is made of nearly $isolated$ CoO$_4$ tetrahedrons with large distance ($\sim 3$ \AA) between them, without any corner-, edge-, or face-sharing coordination~\cite{mattausch1971kenntnis,boulahya2000comparative}. In addition, the three-dimensional (3D) network of nearest-neighbors Co-Co bonds ($\sim 5$ \AA) forms a distorted triangular lattice with high geometric frustration, as shown in Fig.~\ref{mag}. Remarkably, an exotic zigzag AFM order, with the AFM transition temperature $T_{\rm N} \sim 23-26$~K, was reported by neutron scattering experiments and X-ray-diffraction measurements, despite the large distances between adjacent cobalts and the geometric frustration~\cite{jin2006ba,boulahya2006structural,zhang2019anomalous}. Experiments also revealed that an uniaxial magnetoelastic coupling occurs along the $a$ direction~\cite{zhang2019anomalous}.

Previous theoretical studies provided qualitative explanations about the spin exchange interactions, which was assumed to originate from the ``super-super-exchange'' mechanism between CoO$_4$ clusters (namely involving two oxygens as bridge between TM atoms), leading to a BCO zigzag magnetic structure~\cite{koo2006spin}. Recently, another theoretical work suggested, however, that a simplified super-super-exchange is inappropriate for the description of the BCO system, because the oxygen magnetic moment was also considered to provide a substantial contribution to the total local magnetic moments of each CoO$_4$ ionic cluster~\cite{zhang2020magnetic}. Furthermore, their density functional theory (DFT) calculations also unveiled another interesting finding: the spin density exhibits a node between O and Co atoms, and a similar phenomenon was also found in SrRuO$_3$~\cite{zhang2020magnetic,kunkemoller2019magnetization}.

Considering these interesting developments, both in experimental and theoretical efforts, several intriguing questions require a more intuitive theoretical explanation, to gauge to what extend these findings are unique to BCO or whether they could be present in many other materials. {\it First}, what is the driving force of the magnetic polarization on oxygens? Does it really challenge the standard magnetic model ideas for understanding the spin exchange interactions? {\it Second}, what is the physical explanation of the presence of nodes in the spin density along the Co-O bond? {\it Third}, in BCO, why the long-range zigzag spin order can be stabilized in such a highly geometrically frustrated lattice? What is the role of anisotropic magnetic interactions?

To answer these questions, here using simple toy model calculations with the aid of DFT, we carried out systematic studies to provide intuitive explanations of all the unusual properties of the BCO system. First of all, our results based on a simple model provide an interpretation for the presence of nodes in the spin density along the Co-O pathway based on the role of {\it antibonding} molecular levels within the CoO$_4$ cluster ion. In addition, our model calculation explains the unexpected net magnetization present at the ligands. We conclude that this phenomenon (oxygen ion polarization) should be far more general than previously anticipated. We provide a clear rule (as in a theorem) for when oxygen polarization should be detected in a generic oxide
based on whether the overall magnetic order is FM or AFM, and in the latter depending on the number of ligands (even or odd) between TM atoms. Finally, simplified one-orbital Hubbard
and two-orbital double-exchange models (the latter in the Appendix) are used to investigate the formation of the long-range zigzag magnetic order and reveal the importance of the easy-axis anisotropies on triangular lattices to alter the balance between the canonical 120$^{\circ}$ antiferromagnetic order expected in directional isotropic systems vs. the zigzag order found in BCO and in our calculations. The important issues related to magnetoelastic effects in BCO also unveiled in recent work~\cite{zhang2020magnetic} would require incorporating the lattice in our theory effort, and such studies will be carried out in the future.

\section{Node in the spin density along C\lowercase{o}-O bonds in BCO}

In this section, we will provide an explanation for the discovery of zeros in the spin density along the Co-O directions. First, their existence will be confirmed via DFT calculations, finding agreement with results presented before in Ref.~\cite{zhang2020magnetic}. Second, an intuitive explanation will be provided using a simplified model Hamiltonian and relying on antibonding molecular orbitals.

\subsection{DFT spin density along Co-O}
\begin{figure}
\centering
\includegraphics[width=0.48\textwidth]{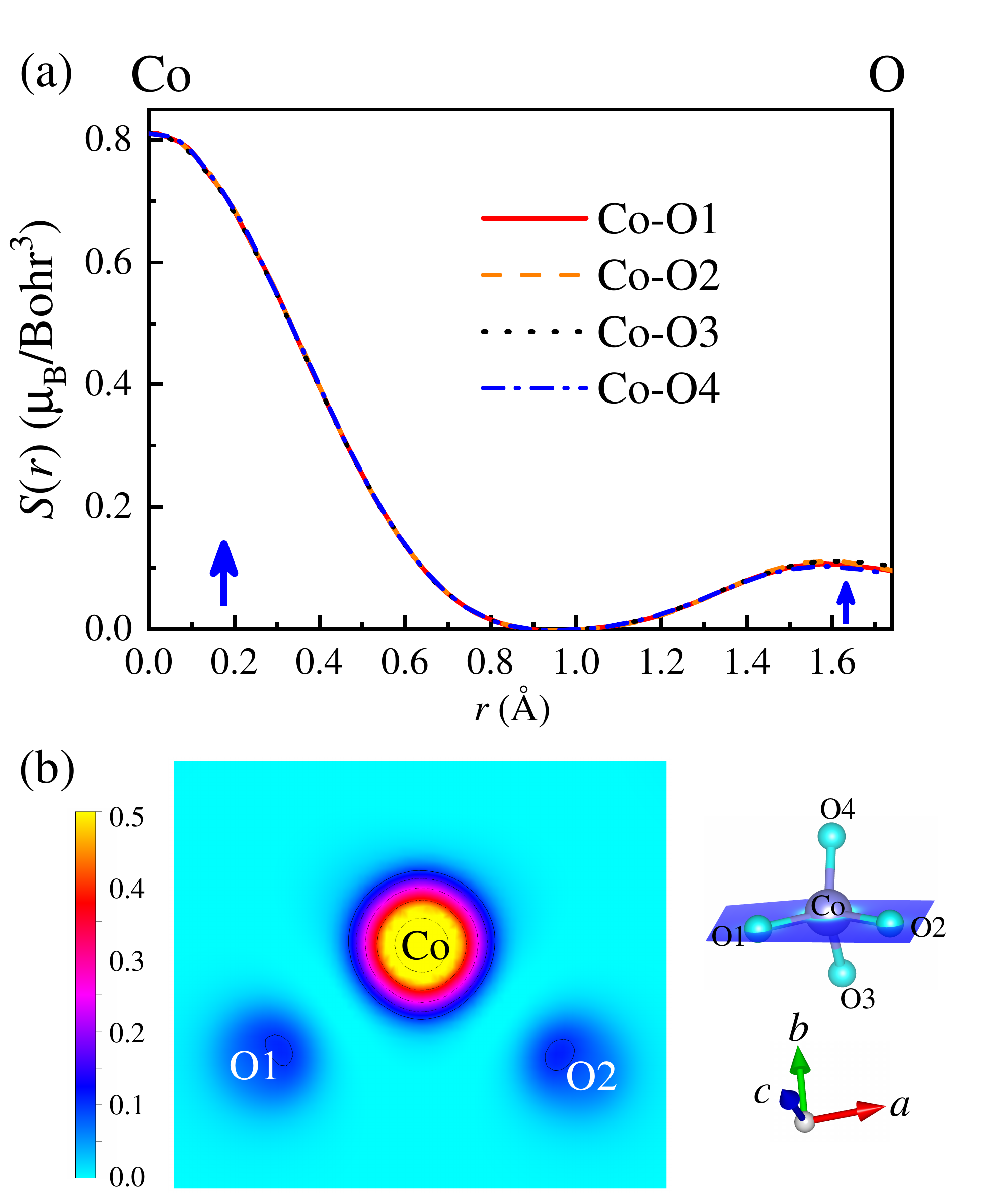}
\caption{(a) Line profiles of DFT spin density $S$($r$) vs. distance $r$ along Co-O bonds for a CoO$_4$ tetrahedron. The four Co-O bonds provide virtually the same result. (b) Contour plot of the spin density projected on the Co-O1-O2 plane. The small illustration on the right shows a sketch of a CoO$_4$ tetrahedron and the corresponding projected plane that we used, in blue.}
\label{dft_CoO_spin_den}
\end{figure}

In the DFT portion of this project, calculations were carried out using the Vienna {\it ab initio} Simulation Package (VASP) code~\cite{Kresse:Prb99,Blochl:Prb2}. The revised Perdew-Burke-Ernzerhof exchange-correlation density functional (PBEsol) was adopted for the exchange correlation potential, as implemented in VASP~\cite{Perdew:Prl,Perdew:Prl08}. The total energy convergence criterion was set to be $10^{-5}$ eV during the self-consistent calculation and the cutoff energy used for the plane-wave basis set was $550$ eV. All calculations were performed with the experimental crystal structure fixed, i.e. without atomistic relaxation. For the non-magnetic phase, the standard primitive unit cell was used and the corresponding $k$-mesh employed was $7 \times 5 \times 3$. From the {\it ab initio} ground-state wave function, the maximally localized Wannier functions~\cite{marzari1997maximally} were constructed using the WANNIER90 code~\cite{mostofi2008wannier90} to extract the crystal field splitting parameters. As for the magnetic phases, we considered the spin-polarized version of the generalized gradient approximation (GGA) potential, which already accounts for the exchange interaction. Thus, an additional $U$ parameter was not included in our calculation. This might underestimate the band gap but still captures the main physics, as discussed in the rest of the text. Because a large magnetic unit cell ($2a \times 1b \times 2c$) was adopted for the magnetic phases, the corresponding $k$-mesh employed was reduced to $4 \times 6 \times 2$.

\begin{figure}
\centering
\includegraphics[width=0.48\textwidth]{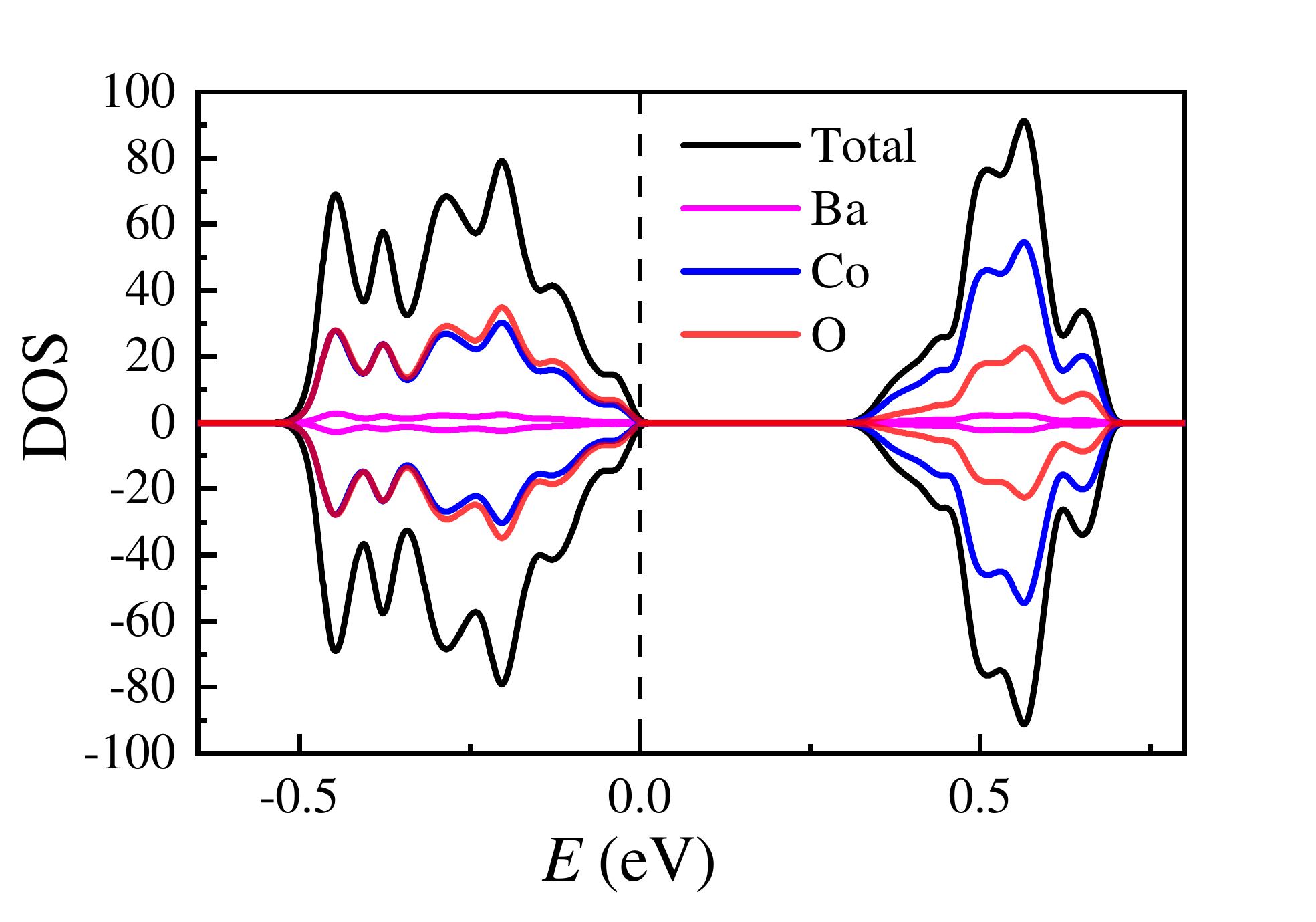}
\caption{The density of states (DOS) of the simplified collinear AFM0 state, from our DFT calculations.}
\label{dos_afm0}
\end{figure}

According to experiments~\cite{zhang2019anomalous}, the Co spins
form a zigzag-type FM configuration along the $b$ axis,
while it is AFM along the $a$ and $c$ axes. The easy-axis magnetization is primarily pointing along the $a$ direction and the spin is slightly non-collinear~\cite{zhang2019anomalous}. For simplicity, Fig.~\ref{mag}~(a) displays the closest {\it collinear} AFM0 state that we constructed. For a better perspective when focusing on the $ab$ plane, the two dimensional (2D) spin order is displayed in Fig.~\ref{mag}~(b). It exhibits a state with a zigzag intra-chain FM pattern along the $b$-axis, with an overall AFM order between these chains.

In Fig.~\ref{dft_CoO_spin_den}~(a), the spin density line profiles along the four Co-O bonds within a CoO$_4$ tetrahedron are displayed. It is clear that the spin density is almost isotropic for the four Co-O bonds. Moreover, an obvious zero, or node, can be found along the path between Co and O. In addition, remarkably, even though the initial magnetic moments on oxygens are set to be zero, a net magnetic moment spontaneously develops on the oxygens and in the same direction as that of Co. The magnetic moments within the default Wigner-Seitz radii of atomic spheres on Co and O atoms are $2.925$ and $0.357$ $\mu_{\rm B}$ per site, respectively. Figure ~\ref{dft_CoO_spin_den}~(b) presents the corresponding 2D contour plot of the spin density projected on the Co-O1-O2 plane.

As shown in Fig.~\ref{dos_afm0}, the corresponding density of states (DOS) around the Fermi level for the AFM0 state has a gap $\sim 0.32$ eV.  Both valence and conduction band are mainly contributed from the Co($3d$) states and the O($2p$) states, while the contribution from Ba is negligible. Obviously, the Co($3d$) states with the O($2p$) states are heavily hybridized particularly for the valence band, which is responsible for the considerable magnetic moment on the O atoms, as shown later in this publication.

In summary, our DFT calculations reproduced the results from a previous study~\cite{zhang2020magnetic}, including the large magnetization on oxygen and the nodes present in the spin density along Co-O bonds.

\subsection{\label{subsec:node}Model calculation}

To understand intuitively the experimental and DFT results regarding the presence of a net magnetization on oxygens and the presence of nodes in the spin density along the Co-O bond, a toy model is here employed that captures the essence of the physics. Understanding intuitively the origin of these effects is important because it will
allow us later in the manuscript to generalize to an arbitrary TM ion with an arbitrary
number of oxygens linking these ions.

As a first step, consider for simplicity a Co atom with 5 active $3d$ orbitals. Co is placed at the origin of coordinates ${\bf r_{Co}}$=(0, 0, 0). The O atom with 3 active $2p$ orbitals is located at the typical distance Co-O for AFM Co-O-O-Co bonds in BCO, namely ${\bf r_O}$=(1.77~\AA, 0, 0). To focus on the essence of the physics, only the active electrons will be considered in our simple models: specifically, 7 $3d$ electrons for each Co$^{2+}$ ion (if S=3/2 is assumed for the Co-ion spin) and 6 $2p$ electrons for the full-shell O$^{2-}$. For the Co-O bond, this gives a total of 13 electrons, for the Co-O-Co bond 20 electrons, and for the Co-O-O-Co bond 26 electrons. Moreover, for the positive charge of the O ion we assume +4 (i.e. the $1s$ and $2s$ fully-occupied orbitals screen the central charge) while for Co we assume a central charge +9 for the same reason (i.e. the fully occupied shells $n=1,2,3$ screen the central charge). This affects the size of the wave functions we employ. Using different effective central charges, within reason, lead to the same qualitative results.

The simple Hamiltonian used here is given by
\begin{equation}
H=\sum_{\sigma}H_{\sigma}
\label{h}
\end{equation}
\noindent where $\sigma=\pm 1$ are the $z$-projections of the spin. The Hamiltonian of each spin sector
has a simple tight-binding hopping term, a Hund coupling at Co aligning the $3d$ spins
(playing in practice the role of an external field), as well as crystal-field energy shifts
between the O and Co orbitals, as it is usual in charge transfer compounds, with values deduced from DFT. Specifically,

\begin{equation}
\begin{aligned}
H_{\sigma}&=\sum_{\gamma,\tilde{\gamma}}t_{\gamma\tilde{\gamma}}(d^{\dagger}_{\gamma,\sigma}p_{\tilde{\gamma}\sigma} + h.c.) \\
&+ \sum_{\gamma} (\sigma J_H + \Delta_{\gamma}) n_{\gamma,\sigma} +\sum_{\tilde{\gamma}} \Delta_{\tilde{\gamma}} n_{\tilde{\gamma},\sigma}
\end{aligned}
\label{htb}
\end{equation}
\noindent where $d^{\dagger}_{\gamma\sigma}$ creates an electron with $z$-axis spin projection $\sigma$ at the Co($3d$) orbitals $\gamma= x^2-y^2, 3z^2-r^2, xy,xz,yz$ while $p^{\dagger}_{\tilde{\gamma}\sigma}$ creates an electron with spin $\sigma$ at the O($2p$) orbitals $\tilde{\gamma}=x,y,z$. The hopping amplitudes $t_{\gamma\tilde{\gamma}}$ correspond to the hybridizations between nearest-neighbors Co and O atoms. Here $n_{\gamma,\sigma}$ ($n_{\tilde{\gamma},\sigma}$) is the number operator for $3d$ ($2p$) electrons with spin $\sigma$.
$J_H$ is the Hund's coupling (or could be considered the effective field created by the rest of the long-ranged order magnetic state in the crystal), while
$\Delta_{\gamma}$ and $\Delta_{\tilde{\gamma}}$ are the on-site energies at the Co and O sites, respectively.

\begin{figure}
\centering
\includegraphics[width=0.48\textwidth]{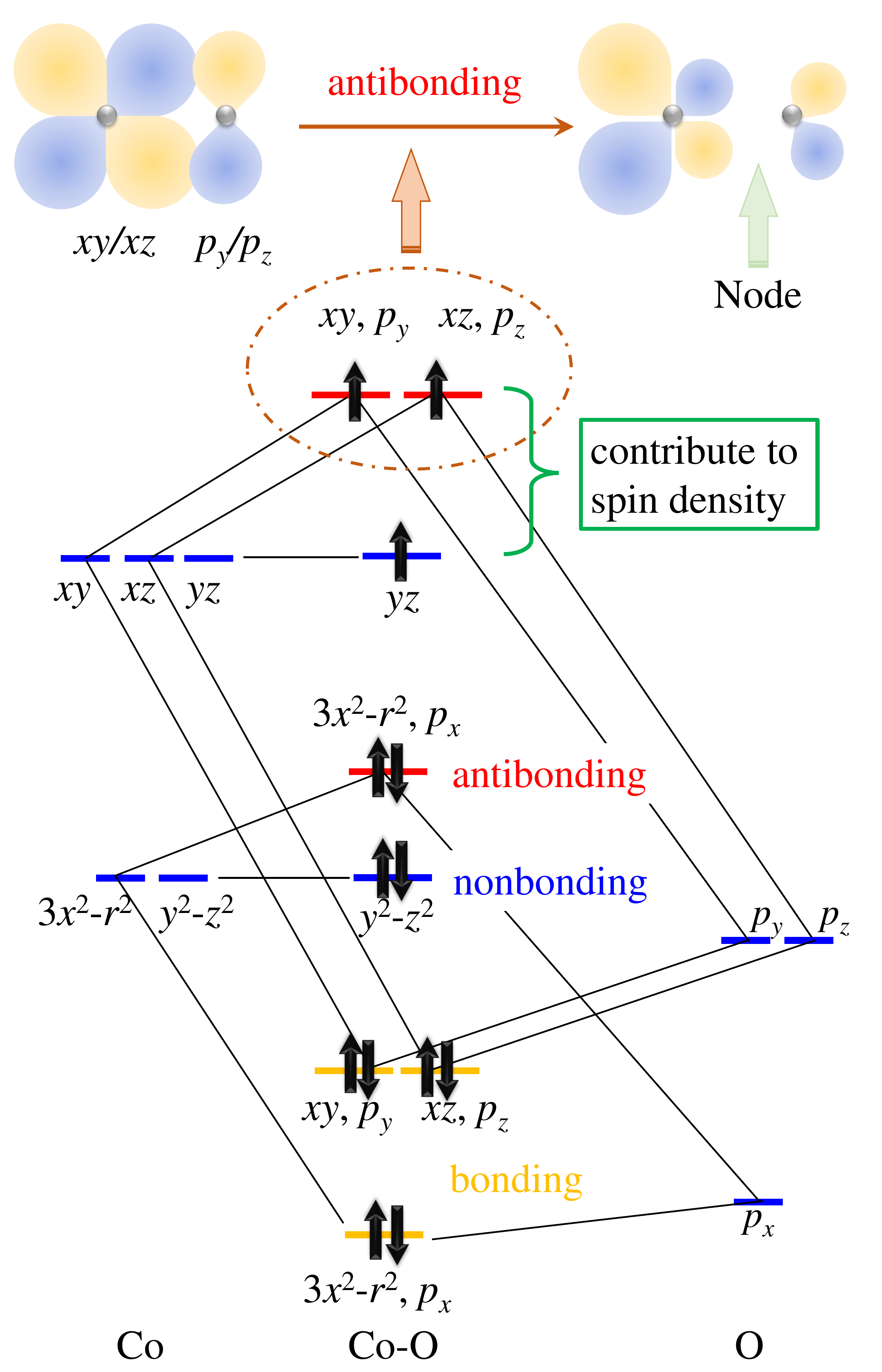}
\caption{Resulting energy level diagram of the molecular orbitals for the Co-O bond, arising from our model calculation. Left: the five $3d$ orbitals of Co with the tetrahedral crystal-field splitting. Right: the three $2p$ orbitals of O, with the $x$-axis located along the Co-O bond direction. Middle: energy level diagram for the Co-O bond molecular orbitals at $J_H=0$. The total population of electrons considered is 8 electrons with spin up and 5 electrons with spin down to fill the energy levels, as shown by the thick arrows. Bonding, nonbonding, and antibonding levels are marked in orange, blue, and red, respectively. Only the singly occupied states make contributions to the spin density. The top panel provides an intuitive view of the formation of the antibonding state of $xy,p_y$ and $xz,p_z$. The node is marked by the green arrow.}
\label{en_le}
\end{figure}

The non-zero hoppings obtained from the Slater-Koster approach can be shown to be $t_{x^2-y^2,x}=\sqrt{3}(pd\sigma)/2-(pd\pi)$, $t_{3z^2-r^2,x}=-(pd\sigma)/2$, $t_{xy,y}=(pd\pi)$, and $t_{xz,z}=(pd\pi)$. For simplicity, we set $(pd\sigma)=1$ and $(pd\pi)=0$, but we confirmed that our
results shown below are qualitatively the same in a range of these couplings (namely as long as the molecular orbitals remain in the same order as in Fig.~\ref{en_le}, to be discussed below). Here we use the hopping amplitudes based on the Slater-Koster approach, instead of those from the Wannier90 fitting, because
Slater-Koster hoppings are generic, simpler to understand, and applicable to any system. The on-site energies, i.e. crystal field splitting, from DFT using Wannier90 are $\Delta_{x^2-y^2}=\Delta_{3z^2-r^2}=0$ (taken as reference energy) and $\Delta_{xy}=\Delta_{xz}=\Delta_{yz}=0.42$ eV for cobalt, consistent with the tetrahedral Co environment of the (CoO$_4$)$^{4-}$ cluster, while for oxygen $\Delta_y=\Delta_z=-0.27$ eV, and $\Delta_x=-1.09$ eV.

The Hund's coupling was set to two values $J_H=0$ and $-0.5$~eV and the results are qualitatively the same. There is no need to fine tune this coupling for the
conclusions found here. To represent a possible total spin S=3/2 at Co, as found in experiments, and considering the usual closed-shell in O$^{2-}$, we have
populated the Co-O bond system with 8 electrons with spin up and 5 electrons with spin down. We have verified that the case Co S=5/2 (with 8 spins up and 3 spins down)
leads to conclusions similar to those discussed below.

\begin{figure}
\centering
\includegraphics[width=0.48\textwidth]{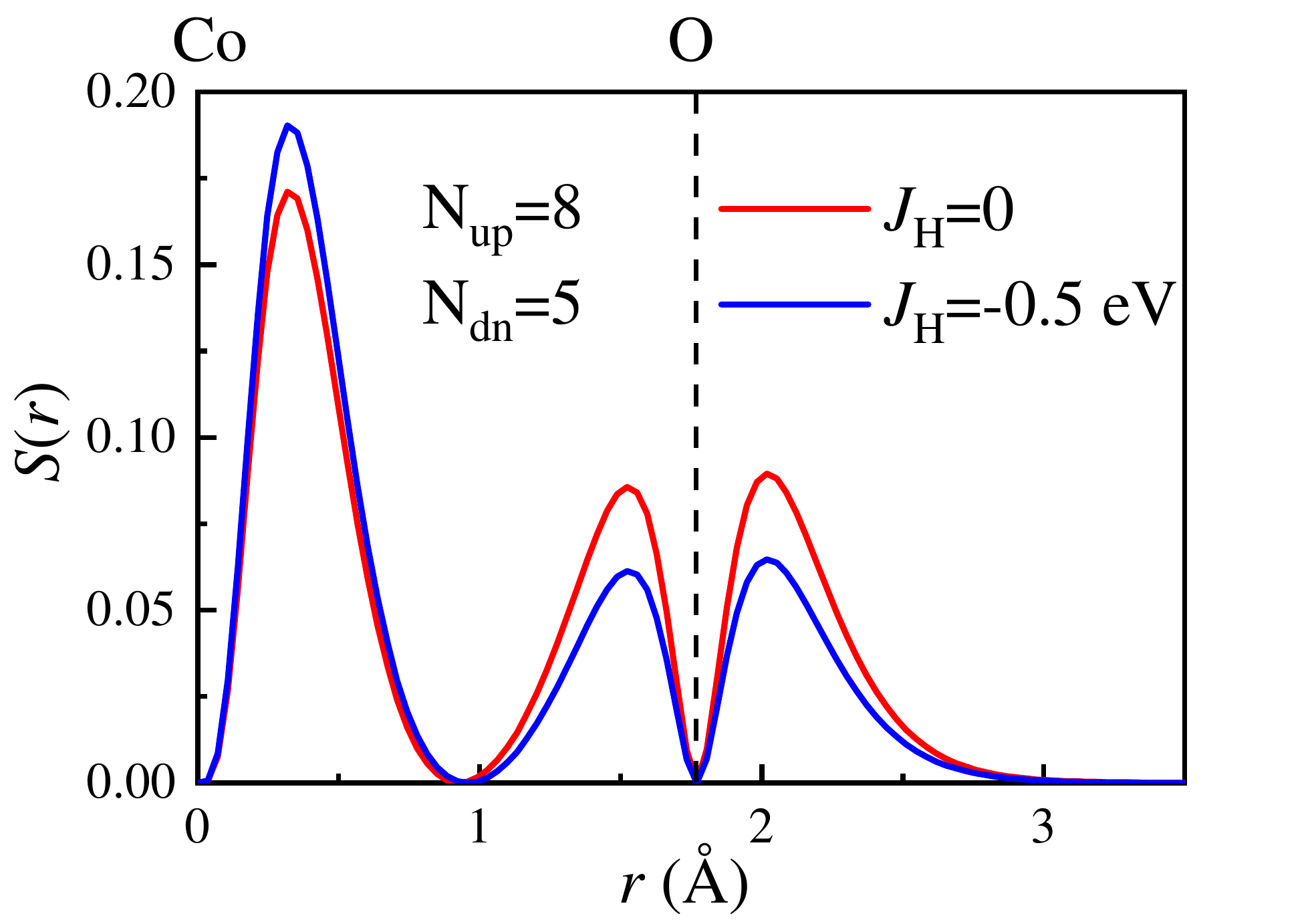}
\caption{Spin density $S$($r$) vs. distance $r$ between a Co at the origin and an O at $r$=(1.77~\AA, 0, 0), the typical Co-O distance in BCO for AFM Co-O-O-Co bonds (here first we focus on one Co-O bond for simplicity). The location of O is marked by the dashed vertical line. The calculation is exact and it includes the 5 $3d$ orbitals at Co and the 3 $2p$ orbitals at O. For this single Co-O bond, 13 electrons were considered (8 with spin up and 5 with spin down). Results are shown for two values of the Hund coupling $J_{H}$. Because the population of electrons already is unbalanced towards spin up, the role of $J_{H}$ is not crucial and even zero Hund coupling leads to the same qualitative physics as a finite Hund coupling. The zero near the center is our main physical focus, while the zeros at the Co and O locations are artificial and caused by the reduced basis used, involving orbitals with nonzero  angular momentum.}
\label{spin_den}
\end{figure}

In Fig.~\ref{spin_den} we show results after diagonalizing the above-described Hamiltonian. The spin density is shown vs. position. There are three zeros along the $x$-axis.
Those at the precise locations of Co and O are trivial: they are induced by our use of a limited set of orbitals all with nonzero angular momentum $l$, which naturally
vanish when their radial coordinate vanishes. The nontrivial zero of our focus is approximately in the middle and in excellent qualitative agreement with previous~\cite{zhang2020magnetic}, as well as our own, DFT results.

To understand intuitively what causes the nontrivial node in the spin density, consider the molecular orbitals resulting from diagonalizing the above-described Hamiltonian. A typical example is shown in Fig.~\ref{en_le} for $J_H=0$. The orientation of the orbitals, their associated wave function signs, and concomitant overlaps define the typical bonding, nonbonding, and antibonding characteristics of these molecular orbitals. The diagonalization establishes that the lowest-energy state is bonding, involving the O $p_x$ and Co $3x^2 - r^2$ orbitals ($3x^2 - r^2$ arises from a linear combination of the standard $3z^2 - r^2$ and $x^2-y^2$ orbitals). The overlap of these orbitals is the largest and, thus, leads to the largest hopping amplitude in the model. The corresponding bonding and antibonding combinations are shown in Fig.~\ref{en_le} and for the population of 13 electrons, they are both doubly occupied. From the bottom, the next levels also correspond to bonding molecular orbitals made of the specific orbitals indicated in the figure i.e. $xy,p_y$ and $xz,p_z$ in our compact notation. They are doubly occupied. However, the antibonding partners have the highest energy among the molecular orbitals, and for the number of spins considered here, only spins up are placed in these antibonding states. Finally, there are also two nonbonding levels, one doubly occupied and the other singly occupied.

After this careful analysis, Fig.~\ref{en_le} then contains the intuitive explanation for the zero or node in the spin density: all the singly occupied levels are either antibonding or nonbonding.
Due to the change in the sign of the wave function between Co and O for the antibonding molecular orbitals, the highest-energy antibonding wave function (singly occupied with a spin up)  generates a zero when the spin density is calculated with the corresponding probability densities (wave function absolute value squared) of each molecular orbital, as sketched in the top panel of Fig.~\ref{en_le}. The singly-occupied nonbonding state does not contribute along the $x$-axis either. For all the doubly-occupied states, the spins up and down contributions cancel out for the spin density. Thus, in summary the spin density develops a zero approximately in the middle between Co and O, as found in DFT, because of the dominance of antibonding molecular orbitals for realistic electronic populations of the Co-O bond.

\section{Polarized Oxygen}

Experimental results for SrRuO$_3$~\cite{kunkemoller2019magnetization} showed that oxygens carry a finite magnetic moment when the Ru spins are in a ferromagnetic state. The recent efforts in Ba$_2$CoO$_4$~\cite{zhang2020magnetic} have reached similar conclusions using DFT, namely the presence of a magnetic moment at the oxygens, but now in a global long-range zigzag antiferromagnetic state (discussed more extensively in the next section) instead of the FM state of SrRuO$_3$.
{\it A priori} finding a magnetic moment at the oxygens is counterintuitive compared with the expected full-shell electronic structure of the O$^{2-}$ ion, with all active $2p$ orbitals doubly occupied. However, as explained in the case of the lone Co-O pair when we discussed the zero in the spin density in Subsec.~\ref{subsec:node}, we also found a nonzero magnetic moment at the oxygen merely as a product of electronic itineracy: the spins {\it down} jump from O to Co to optimize the kinetic energy, thus leaving behind a positive net moment at the O and reducing the moment at Co. In this section, we will generalize this last conclusion to other situations, both for FM and AFM states, now involving pairs of Co spins.

\subsection{One oxygen as bridge}
\begin{figure}
\centering
\includegraphics[width=0.48\textwidth]{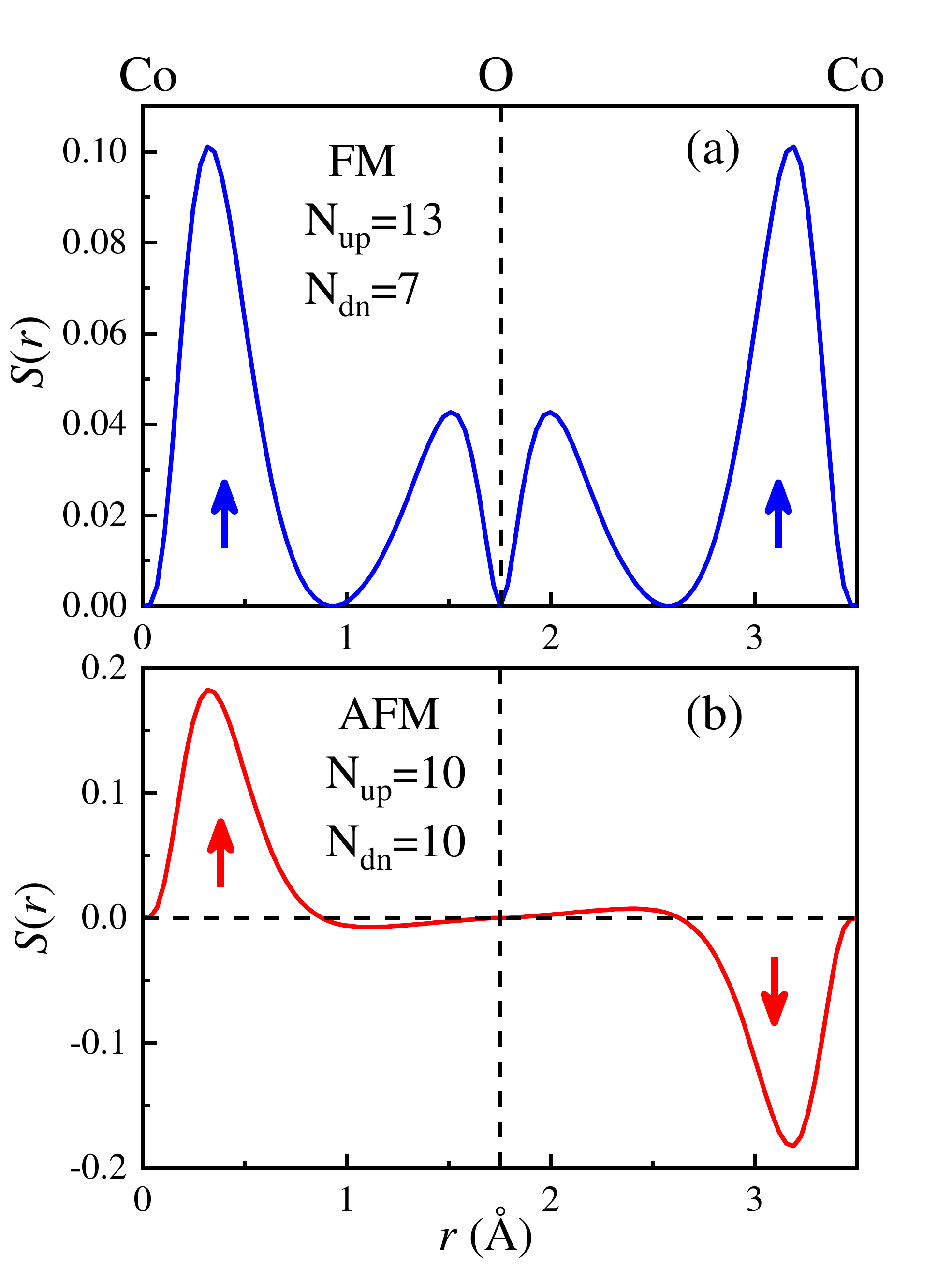}
\caption{Spin density $S$($r$) vs. distance $r$ for a Co-O-Co cluster, using our simplified model calculation.
The O is at $r$=(1.77~\AA, 0, 0), as in the previous section. The Co's are located at the origin and at $r$=(3.54~\AA, 0, 0). Here we simply doubled the Co-O of the left to create a symmetric bond with O exactly at the center,
as example of the physics we wish to explain. Other Co-O distances lead to the same conclusions as long as the bond
is left-right symmetric.
The location of O is marked by a dashed vertical line. The calculation is exact and includes 10 $3d$ orbitals at the two Co's combined and 3 $2p$ orbitals at the O. (a) FM case, with 13 spin-up electrons and 7 spin-down electrons, such that the entire Co-O-Co structure carries a net magnetization. The Hund coupling is chosen as $J_H=-2$ for both Co's, merely as example. The zeros at exactly Co and O locations are artificial due to the reduced basis used. (b) AFM case, with 10 spin-up electrons and 10 spin-down electrons (for the AFM case the overall net magnetization must vanish). The Hund couplings must have different signs for each Co to simulate the AFM configuration.
Here we select as example $J_H=-1$ for the left Co and $J_H=+1$ for the right one, respectively.}
\label{spin_fm_afm}
\end{figure}
We will consider first a simple extension of the Co-O calculations in Fig.~\ref{spin_den}. We will start with a three-atoms Co-O-Co arrangement. This qualitatively occurs in perovskite cobaltites but not in Ba$_2$CoO$_4$ where two oxygens act as bridge between cobalts. But the Co-O-Co example discussed here provides a clear starting point and the double-O bridge will be addressed next. Moreover, any distance Co-O leads to the same qualitative conclusions. The calculation now uses 5 $3d$ orbitals in the left Co, 3 $2p$ orbitals in the central O, and another 5 $3d$ orbitals in the right Co. The hoppings Co-O and O-Co are the same by symmetry:
for example, the different + and - lobe signs of the central $p$ orbital along the $x$-axis is compensated by choosing $y^2 - x^2$ for the right Co
instead of $x^2-y^2$ so that bonding orbitals have always the lowest energies. The diagonalization is now repeated with the obvious addition of an extra site in the
model described in Subsec.~\ref{subsec:node}, enlarging the Hilbert space. The results are shown in Fig.~\ref{spin_fm_afm}.

\begin{figure}
\centering
\includegraphics[width=0.48\textwidth]{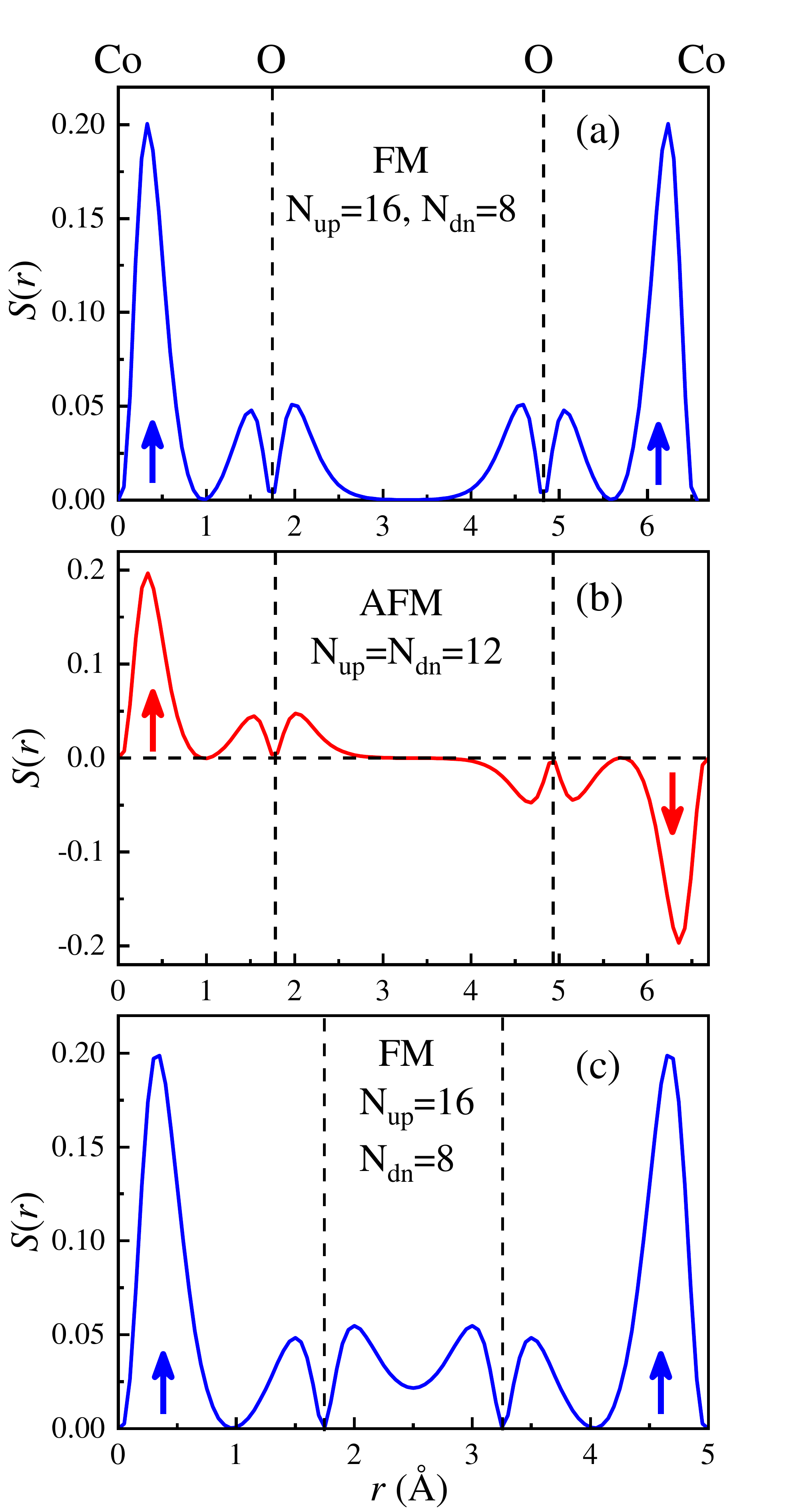}
\caption{Spin density $S$($r$) vs. distance $r$ for a Co-O-O-Co cluster, using our simplified model calculation. The locations of O's are marked by dashed vertical lines. The calculation is exact and it includes 5 $3d$ orbitals at the Co's and 3 $2p$ orbitals at the O's, for a total of 16 orbitals. The Co Hund coupling $J_H=-1$ for all cases, as example. (a) FM case, considering overall 16 electrons with spin up and 10 electrons with spin down. (b) AFM case, considering 13 electrons with spin up and 13 electrons with spin down (the overall Co-O-O-Co must have a zero net magnetization for the AFM case). (c) FM case but with an artificially smaller distance between the O-O bond to enlarge the spin density at the center which was exponentially suppressed in (a), and, thus, invisible to the eye in practice, due to the exponentially localized nature of the wave functions.}
\label{spin_fm_afm_real}
\end{figure}

\begin{figure}
\centering
\includegraphics[width=0.48\textwidth]{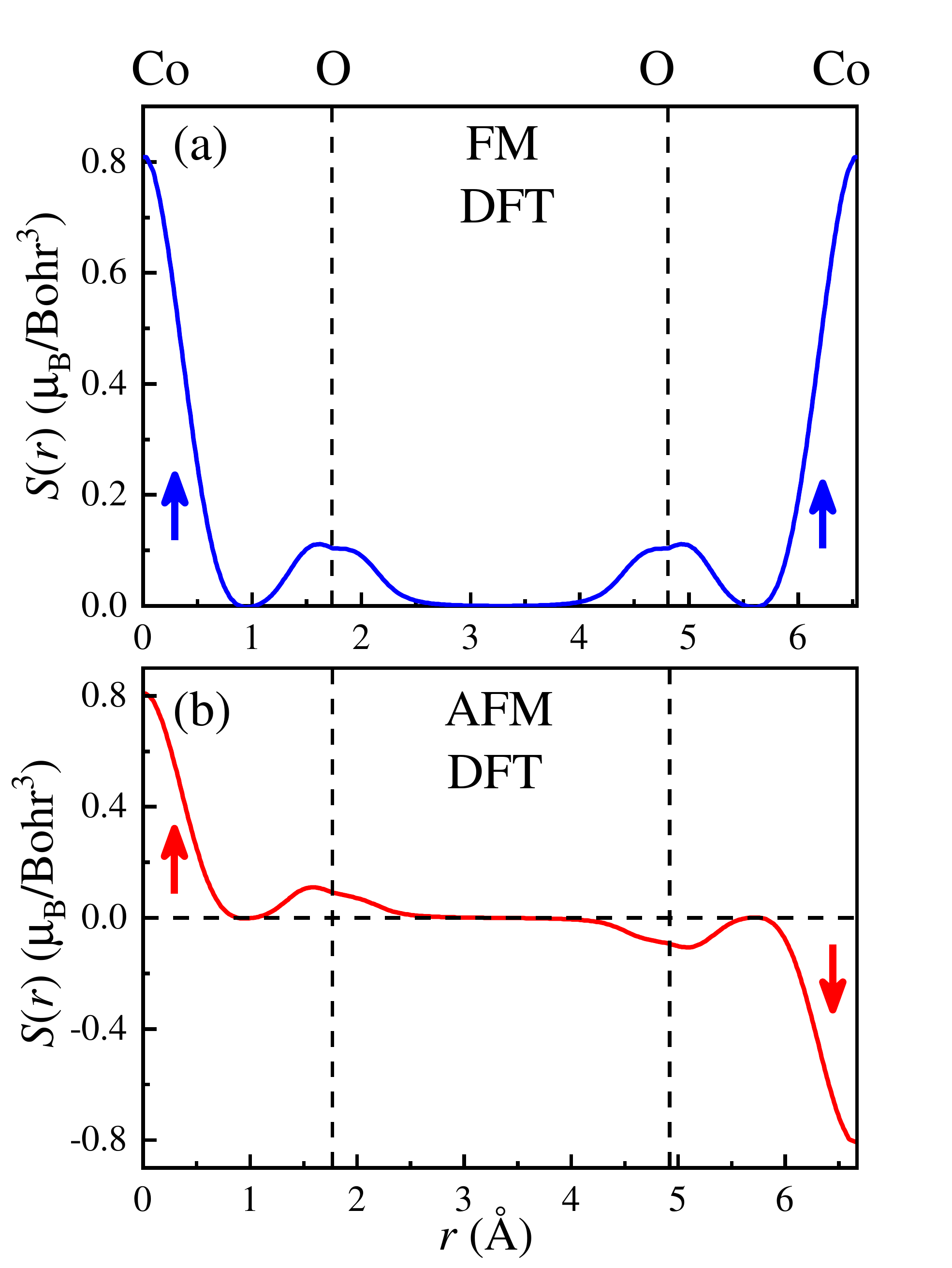}
\caption{Line profiles of DFT spin density $S$($r$) vs. distance $r$ for a Co-O-O-Co path in BCO. (a) FM case, such as along the Co$_{14}$-O-O-Co$_{10}$ path, corresponding to the red path in Fig.~\ref{mag} (b). (b) AFM case, such as along the Co$_{14}$-O-O-Co$_{16}$ path, corresponding to the blue path in Fig.~\ref{mag} (b). The locations of O's are marked by vertical dashed lines. Note that the actual Co$_{14}$-O-O-Co$_{10}$ paths are bended
in Ba$_2$CoO$_4$ because the locations of the Co and O are not in a straight line. These figures show that our simple model and the more sophisticated DFT calculations qualitatively agree.}
\label{dft_spin_fm_afm}
\end{figure}

\begin{figure}
\centering
\includegraphics[width=0.45\textwidth]{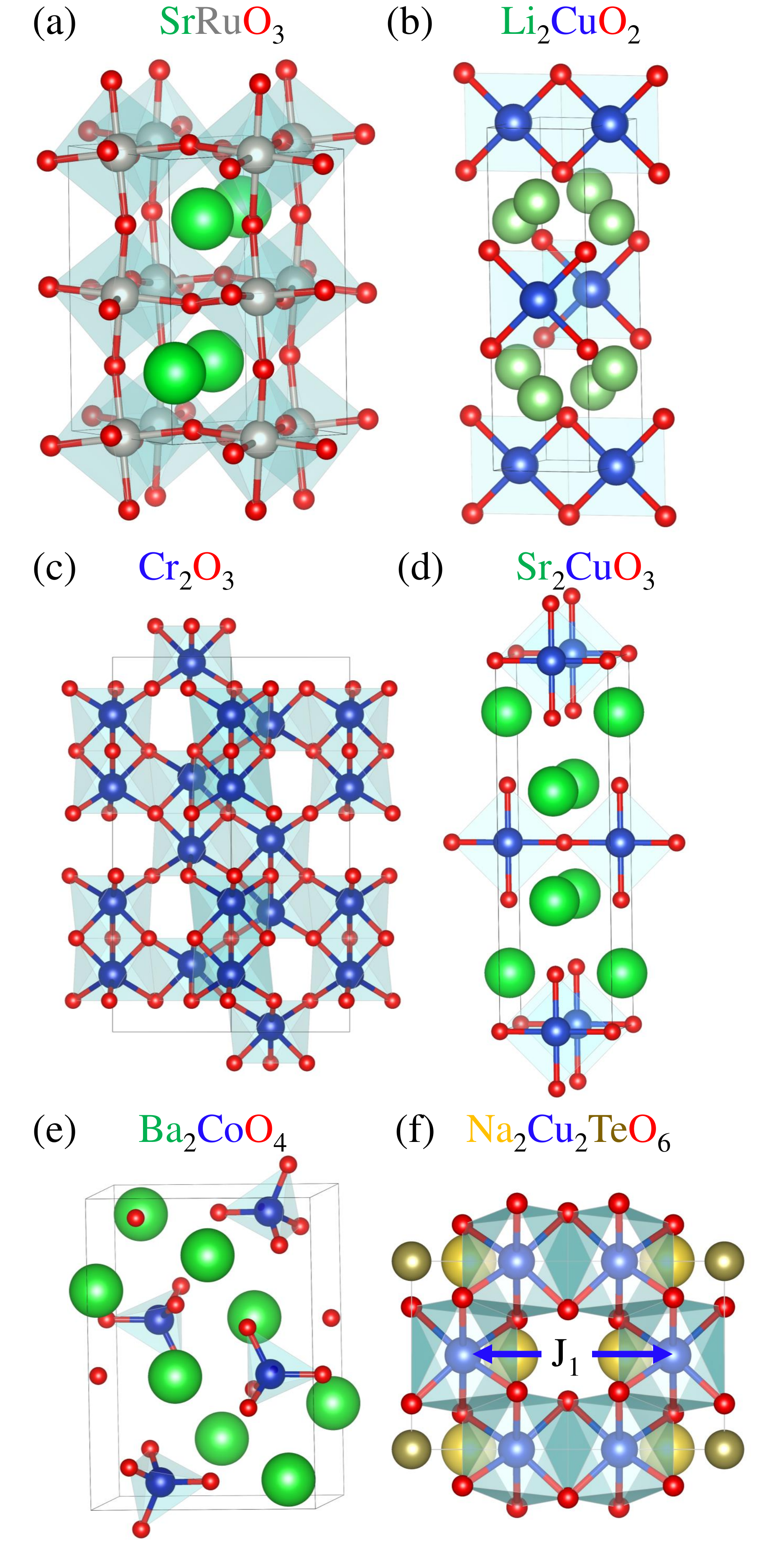}
\caption{Examples of crystal structures of selected real materials where our results are applicable. TM-O polyhedrons are shown in translucent blue. (a)-(d) For the case of one oxygen as bridge. (e)-(f) For the case of two oxygens as bridge. Note that one material might include both the one oxygen path and the two oxygen path. Here we only focus on the most important path.}
\label{material}
\end{figure}

Let us focus first on the Co FM configuration, Fig.~\ref{spin_fm_afm}~(a). The total population of electrons considered here is 13 spins up and 7 spins down, with the two Co spins pointing along the $z$ axis. This is the natural extension of the previous Co-O calculation. Before the hopping terms are included, this crudely corresponds to a total spin S=3/2 at each Co and S=0 at the O. All the parameters in the model Hamiltonian are the same as in Subsec.~\ref{subsec:node}. We show results only for one Hund coupling, but changing $J_H$ only alters the quantitative values of the magnetic moments, but not the qualitative results.  Here the resulting spin density is symmetric with respect to the center. The parallel arrangement of Co spins continue inducing a $finite$ magnetic moment at the central oxygen, due to the electronic mobility contained in the tight-binding hopping. Thus, it is natural that the bridge oxygens, or other ligands, in a FM environment will develop a net nonzero magnetic moment, even if the dominant electronic configuration of oxygen is spinless.

This should occur also in manganites (La,Sr)MnO$_3$ or (La,Ca)MnO$_3$ at hole dopings in their FM state, as well as in any other ferromagnetic state with any kind of full-shell ligand as bridge between spins: the {\it a priori} spinless ligand will always develop a net polarization. Its specific strength may depend on the hopping values, charge transfer gap, and atomic distances, but the ligand magnetic moment will {\it always} be nonzero in a FM state. Generic ideas, such as double exchange that relies on ferromagnetism induced by a large Hund coupling at the TM, are not qualitatively affected by the O magnetization, but our results suggest that O may need to be incorporated for better quantitative predictions. Other possible real materials examples for this case are SrRuO$_3$~\cite{kunkemoller2019magnetization}(Fig.~\ref{material} (a)), yttrium iron garnet~\cite{bonnet1979polarized}, Li$_2$CuO$_2$~\cite{chung2003oxygen} (Fig.~\ref{material} (b)), La$_{0.8}$Sr$_{0.2}$MnO$_3$~\cite{pierre1998polarized}, YTiO$_3$~\cite{kibalin2017spin}, Ca$_{1.5}$Sr$_{0.5}$RuO$_4$ \cite{gukasov2002anomalous}, and Sr$_2$IrO$_4$~\cite{jeong2020magnetization}.

Consider now the case of Fig.~\ref{spin_fm_afm}~(b) with the Co atoms in an AFM configuration. The central O now has a net zero magnetic moment because the (very small) positive magnetization on one side of O exactly cancels the negative magnetization on the other side of O. While in the FM arrangement the Co spins favor the positive direction of the $z$-axis by breaking the symmetry up-down along that axis, in the AFM arrangement the central oxygen is subject to opposite tendencies of equal magnitude. Thus, its net magnetic moment must be zero, as shown in the calculation and in the figure: the additions of the oxygen spin density left and right of its center just cancels. Note that this is different from having zero spin density at all points in the oxygen range, as for O$^{2-}$. Point by point, the spin density is nonzero at the oxygen (although very small). But still the net result is a zero magnetization for the oxygens for the Co-O-Co combination. Thus, in AFM states with one O as bridge, such as in the CuO$_2$ layers of cuprates, experimentally the oxygen should have no net spin unless there is some lattice distortion that breaks the parity symmetry of the bond. However, the spin density $S(r)$ will be small but nonzero, with different signs left and right. These predictions for $S(r)$ in AFM bonds for one O as bridge, or more as discussed below, could be confirmed by polarized neutron diffraction experiment. Possible real materials examples for this case are Cr$_2$O$_3$~\cite{brown2002determination} (Fig.~\ref{material} (c)), and Sr$_2$CuO$_3$~\cite{walters2009effect}(Fig.~\ref{material} (d)).

\subsection{Two oxygens as bridge}

Let us address now the specific case of Ba$_2$CoO$_4$ where the Co spins are bridged by {\it two} oxygens, namely where the AFM order is mediated by super-super-exchange. In our model we can study the two-oxygens bridge employing the same approach used before, but now introducing a third Slater-Koster parameter (pp$\sigma$) for the overlaps between $2p$ orbitals among the two oxygens at the center of the Co-O-O-Co structure. In Fig.~\ref{spin_fm_afm_real} we show results both for the FM and AFM Co configurations.

For the FM case, Fig.~\ref{spin_fm_afm_real} (a), the results confirm and extend the conclusions of the study using Co-O and Co-O-Co: as a consequence of the mobility of electrons via hopping between atoms, the original finite magnetization at the Co sites spreads to the neighboring O sites, creating a finite spin magnetization at the oxygens similarly as found experimentally in SrRuO$_3$. Because of the large distance 3.08 \AA~for FM bonds between the central oxygens in Ba$_2$CoO$_4$ (3.15 \AA~for AFM bonds), the spin density at the center of Fig.~\ref{spin_fm_afm_real} (a) is exponentially suppressed, and the two Co-O portions appear as almost disconnected. But if we artificially reduce their distance, the central spin density can become robust and visible. This is exemplified in Fig.~\ref{spin_fm_afm_real} (c) where the O-O distance was made smaller than Co-O. Possible real materials examples are CuWO$_4$, CuMoO$_4$-III, and Cu(Mo$_{0.25}$W$_{0.75}$)O$_4$ with O-O distance $\sim2.4$ \AA~\cite{koo2001spin}, but our main goal in panel (c) is only illustrative. Figure~\ref{spin_fm_afm_real} (c) simply displays qualitatively the essence of our prediction under general circumstances. All oxygens in between super-super-exchange coupled transition metal spins, when in a FM state, will develop a finite polarization due to the influence of the nearby finite-spin TM ions via the nonzero tunneling probability for electrons with the opposite spin to the orientation of the TM spins (the spins up are frozen by the Pauli's principle for the number of electrons and orbitals used here).

For the AFM case, Fig.~\ref{spin_fm_afm_real} (b), the result is even more interesting. Note that now {\it by mere geometry} each of the two oxygens is either closer to one Co
or to the other Co. Thus, each oxygen is influenced differently by the environment they are immersed in. Thus, as shown in Fig.~\ref{spin_fm_afm_real}(b), in the AFM Co-O-O-Co structure now both oxygens develop a nonzero magnetic moment. Moreover, these moments have different signs. This is contrary to AFM Co-O-Co where the central oxygen had a net zero magnetization, because it is at equal distance from the Co atoms. Our predictions are in excellent agreement with DFT results, as shown in Fig.~\ref{dft_spin_fm_afm}.

Another real material example for having two oxygens as bridge is the Cu-O-O-Cu $J_1$ path in Na$_2$Cu$_2$TeO$_6$ (Fig.~\ref{material} (f)), where we recently also found a small net magnetization at the oxygens $\sim 0.05~\mu_{\rm B}$ for {\it both} the FM or AFM states from DFT calculations~\cite{gao2020weakly}.

Once again, we remind the readers that the zeros at exactly the Co and O positions in the model Fig.~\ref{spin_fm_afm_real} are spurious due to the small subset of orbitals used in the simplistic model. In DFT these spurious zeros are not present because more orbitals, with plane-wave basis sets instead of atomic orbital basis sets, contribute to the spin density while another possible contribution arises from the compensation portion of the pseudo orbitals inside the projector augmented wave spheres. Thus, in DFT only the important nodes between Co and O appear.

\begin{figure}
\centering
\includegraphics[width=0.48\textwidth]{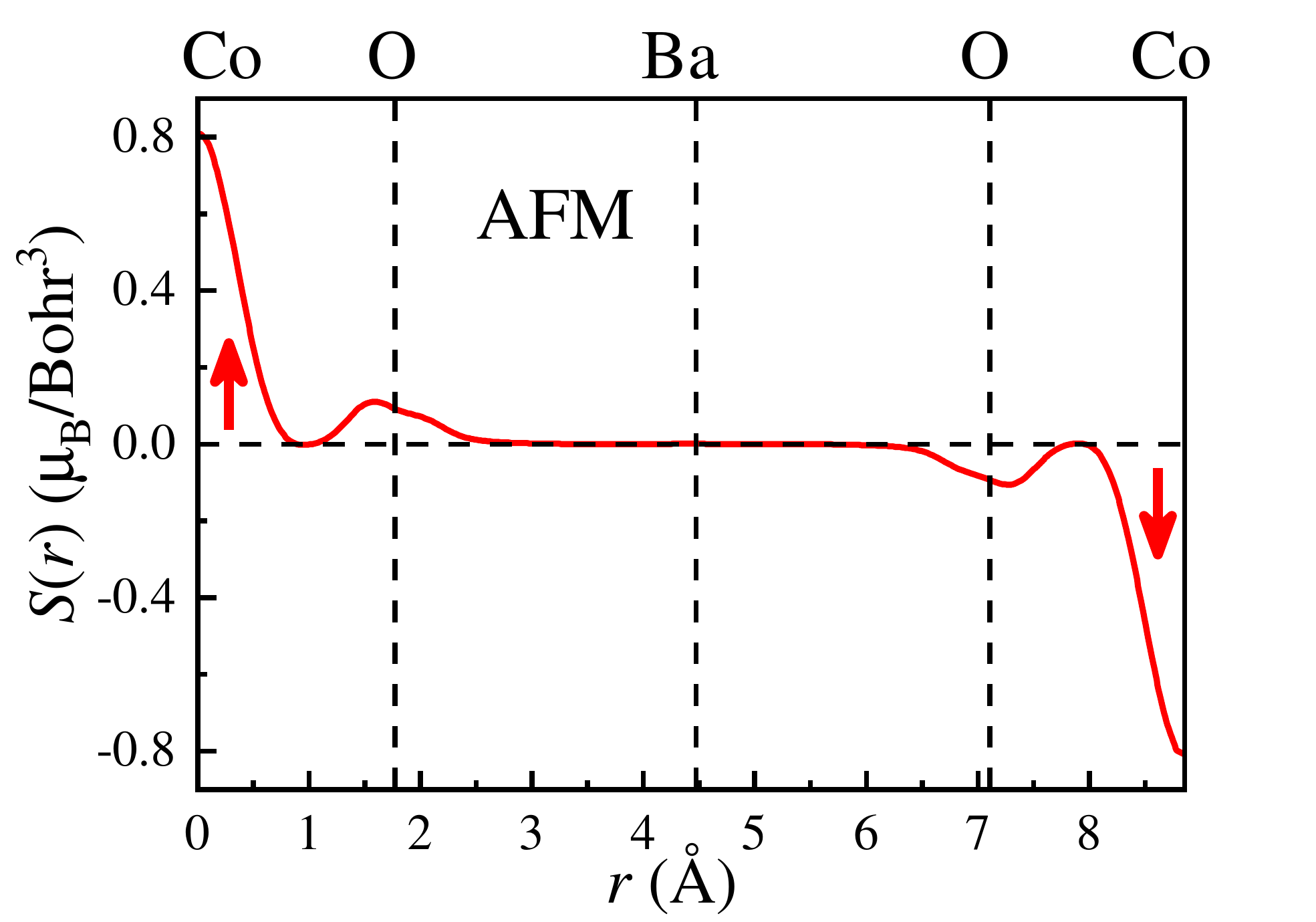}
\caption{Line profiles of DFT spin density $S$($r$) vs. distance $r$ for a Co-O-Ba-O-Co path. Shown is an AFM case, such as along Co$_{14}$-O-Ba-O-Co$_{16}$ in the blue path in Fig.~\ref{mag}. Ba and O's are marked by vertical dashed lines. In BCO, the actual Co$_{14}$-O-Ba-O-Co$_{16}$ path is bended i.e. the locations of the Co, Ba and O atoms are not in a straight line. The main results of this figure is that oxygens have a net spin while Ba does not have a net spin (even its spin density $S(r)$ is exponentially suppressed by the wave functions used). This corresponds to Fig.~\ref{oxy_pol}~(c). Note that in materials with a smaller O-Ba-O distance
the central Ba spin density $S(r)$ would become visible but their net magnetization would still be zero.}
\label{dft_Ba_spin_den}
\end{figure}

\begin{figure*}
\centering
\includegraphics[width=0.92\textwidth]{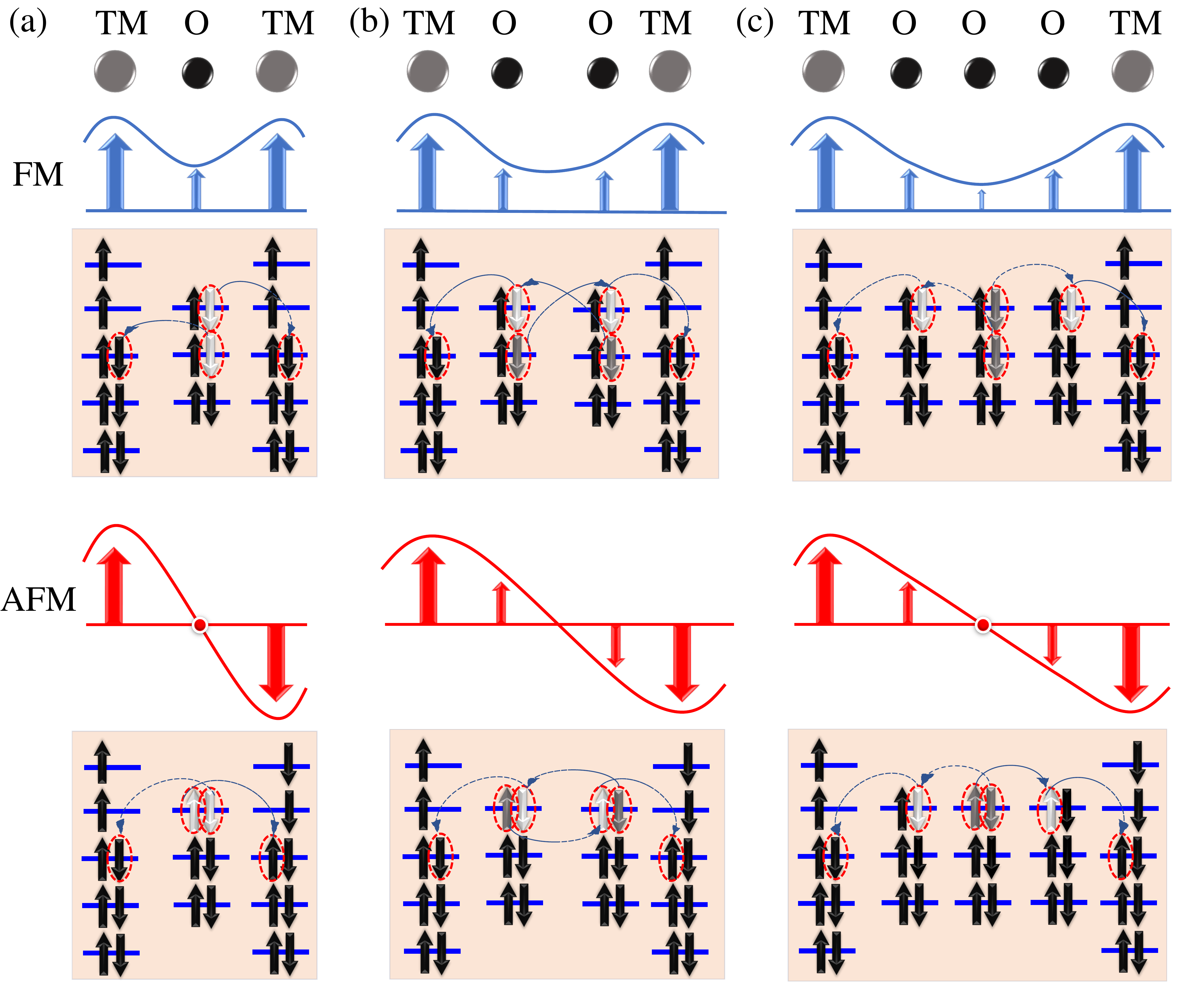}
\caption{Sketches of our general prediction for the net oxygen magnetic polarization corresponding to (a) one, (b) two, and (c) three oxygens (or other ligands) as bridge between two transition metal (TM) elements with a net spin. Both FM and AFM configurations in the TMs are considered here. Illustrations of mechanisms for the development of magnetization for each case are shown below accordingly.}
\label{oxy_pol}
\end{figure*}

\subsection{Three or more oxygens as bridge}
The above described results already provide a natural generalization in case there are materials where the path between transition metal spins contains more oxygens or ligands, such as three. A summary of our conclusions is in Fig.~\ref{oxy_pol} for the cases of 1, 2, and 3 oxygens, and both for the FM and AFM configurations of transition
metal spins. For each case, the pictorial understanding for the formation of a net oxygen magnetic polarization is shown below, respectively. Consider the FM case with only one ligand as an example. Here the electron movement is severely restricted by the Pauli principle. In fact, only the mobility of the electrons with spins down, {\it opposite} to the spins of the single occupied orbitals of the closest TM atoms, is allowed and it develops a net magnetization on oxygens. By contrast, for the AFM case with only one ligand, by symmetry, the mobility of the electrons with spins down and up cancels out and would not develop a net magnetization on oxygen.

For the FM configuration, top sketches, our concrete prediction is that always there will be a net magnetization at the ligands in between transition metals with nonzero spin. The magnitude will decrease towards the center, as shown for the case of three oxygens, and it may become difficult to detect experimentally. But our calculations are intuitively clear, thus we can predict with confidence that with enough experimental accuracy all oxygens will display a net magnetic moment, for both an even or odd number of oxygens, in a FM bond.

For the AFM configuration our general conclusion is more subtle and shown in the lower panels of Fig.~\ref{oxy_pol}. For the case of an {\it even} number of oxygens, as in (b), both oxygens develop a nonzero magnetic moment as in the FM case, and as also exemplified in our study of Co-O-O-Co using the Slater-Koster hoppings. However, for an {\it odd} number of oxygens by symmetry the central one carries an exactly zero total magnetic moment (unless the crystal
is distorted and the left-right parity symmetry of the bond is altered). This conclusion for an odd oxygen number is in Fig.~\ref{oxy_pol}~(a), as in Cu-O-Cu for the cuprates. In Fig.~\ref{oxy_pol}~(c), the central oxygen also has a vanishing magnetic moment, while the other two have a finite magnetic moment. A possible example of the case (c) is LaFeAsO along the $c$-axis, involving an As-O-As bridge. As another possible example for (c), we use the spin density of the Co-O-Ba-O-Co link with Ba instead of O, obtained from DFT, as shown in Fig.~\ref{dft_Ba_spin_den}.

\section{Long-range zigzag spin order}
In this section we will discuss the origin of the exotic zigzag spin pattern
found in Ba$_2$CoO$_4$~\cite{zhang2019anomalous,zhang2020magnetic}.

\subsection{DFT calculation}
\begin{figure}
\centering
\includegraphics[width=0.48\textwidth]{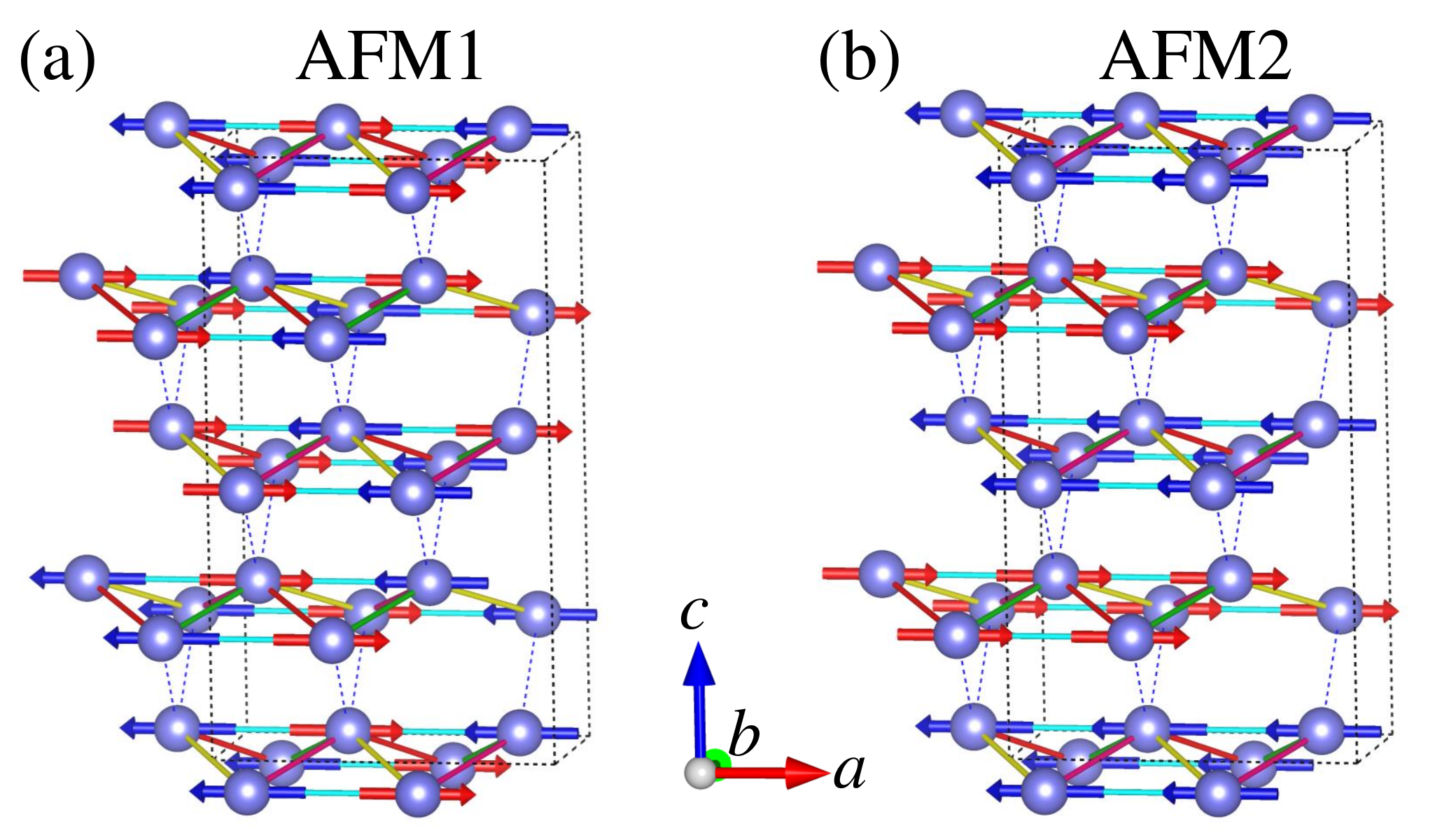}
\caption{Side view of other spin configurations (a) AFM1 and (b) AFM2.}
\label{mag1}
\end{figure}

\begin{table}
\centering
\caption{List of DFT calculated energies of four collinear spin configurations in the same magnetic unit cell. The AFM0 state discussed earlier is the energy reference. All other states studied have higher energy showing that zigzag is the true ground state.}
\begin{tabular*}{0.48\textwidth}{@{\extracolsep{\fill}}lccc}
\hline
\hline
Marks& Configurations & Energy (meV)\\
\hline
AFM0   & $\uparrow\downarrow\downarrow\uparrow\uparrow\downarrow\downarrow\uparrow\downarrow\uparrow\uparrow\downarrow\downarrow\uparrow\uparrow\downarrow$& 0\\
FM & $\uparrow\uparrow\uparrow\uparrow\uparrow\uparrow\uparrow\uparrow\uparrow\uparrow\uparrow\uparrow\uparrow\uparrow\uparrow\uparrow$ & 633\\
AFM1 & $\uparrow\downarrow\downarrow\uparrow\uparrow\downarrow\downarrow\uparrow\uparrow\downarrow\downarrow\uparrow\uparrow\downarrow\downarrow\uparrow$&  68\\
AFM2& $\uparrow\uparrow\uparrow\uparrow\uparrow\uparrow\uparrow\uparrow\downarrow\downarrow\downarrow\downarrow\downarrow\downarrow\downarrow\downarrow$ & 157\\
\hline
\end{tabular*}
\label{table1}
\end{table}

By using the same magnetic unit cell as for the previously-discussed zigzag collinear state AFM0, we also calculated other three different spin configurations (see Fig.~\ref{mag1}) for comparison, as shown in Table ~\ref{table1}, AFM0 is confirmed to be the ground state, as in experiments.
Note that in our DFT calculation, the easy-axis magnetization is not specified to be along a crystallographic direction, i.e. the spin has no preferred direction. This DFT exercise
shows that the zigzag order is robust and our intention in the rest of this section is to provide an intuitive explanation for why this order arises using a simple one-orbital Hubbard model adapted to BCO. In the Appendix, we will also use a two-orbital double-exchange model {\it not} for BCO but in order to illustrate that in a triangular lattice under fairly general circumstances once the canonical large-Hubbard-$U$ 120$^{\circ}$ order is suppressed by strong easy-axis anisotropies, then zigzag patterns emerge.

\subsection{One-orbital Hubbard model with easy-axis anisotropy}
\begin{figure*}[ht]
\centering
\includegraphics[width=0.96\textwidth]{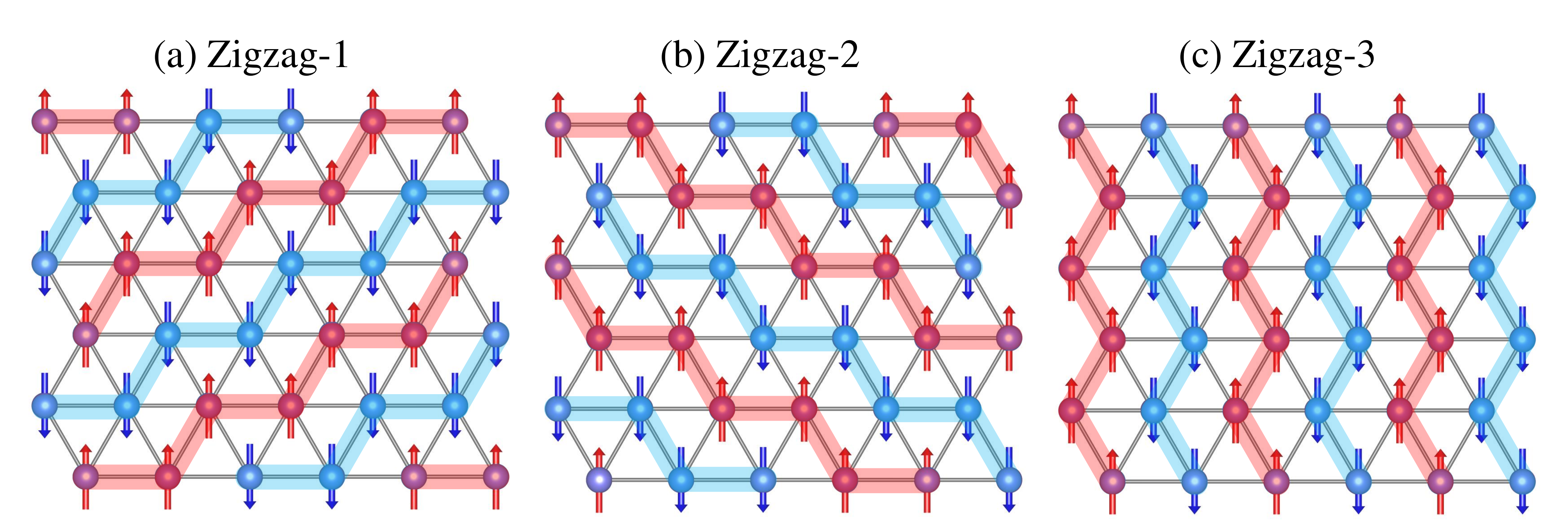}
\caption{The three degenerate zigzag ground states shown using a triangular $6\times6$ cluster as illustration.}
\label{mag_hu}
\end{figure*}

\begin{figure}
\centering
\includegraphics[width=0.48\textwidth]{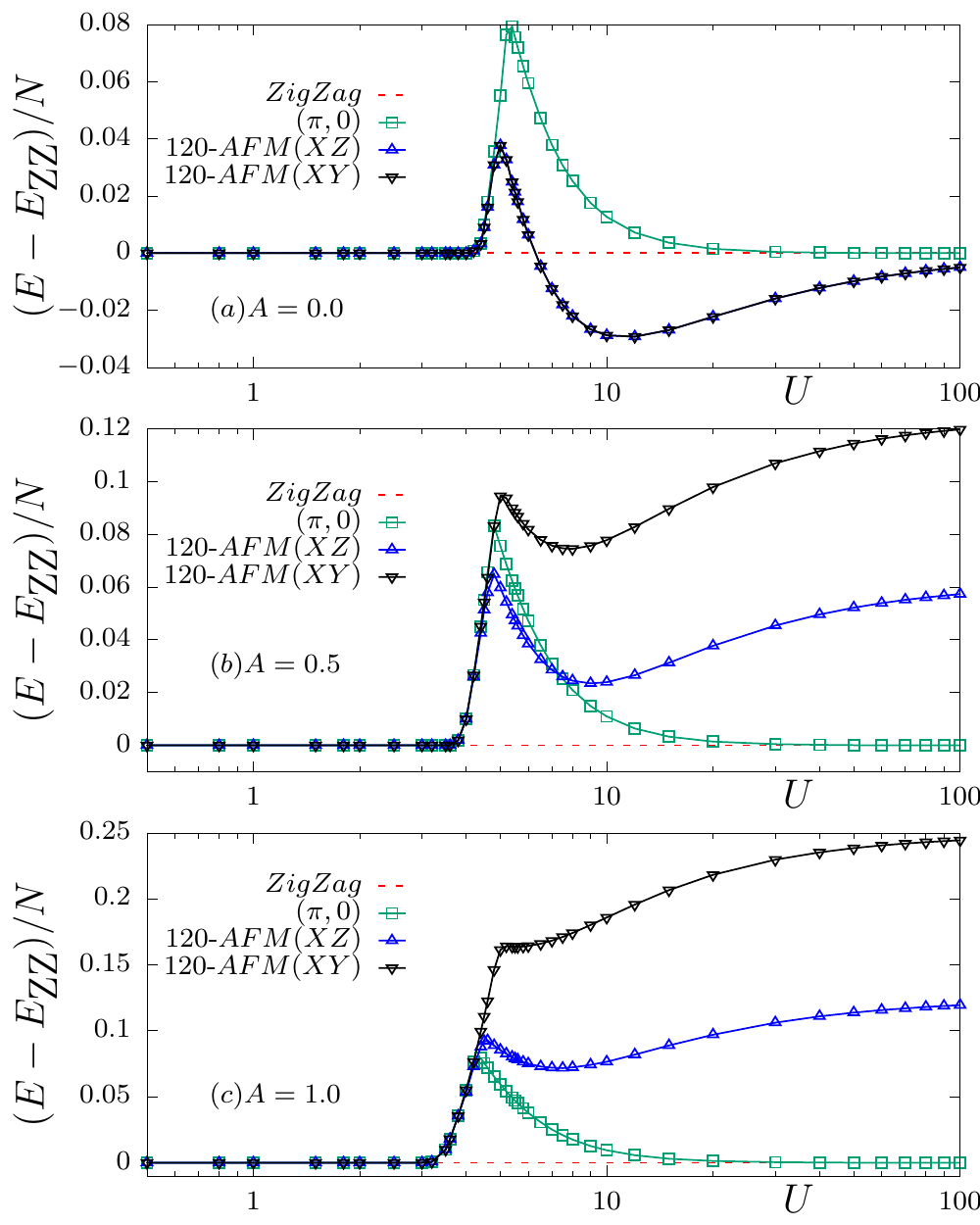}
\caption{The energies per site for various ansatz states with respect to the Zigzag state for anisotropy $A= 0.0$, $0.5$, and $1.0$ in unit of $t$, respectively. System size is fixed to $24 \times 24$, and $N$ is number of sites.}
\label{comparison}
\end{figure}

To better understand the ground state long-range spin order of Ba$_2$CoO$_4$ and the importance of the easy-axis anisotropy, a one-orbital Hubbard model on a triangular lattice was studied \cite{footnote}. The model is defined, in standard notation, as
\begin{equation}
\begin{aligned}
H&= -t\sum_{\langle i,j \rangle,\sigma} (c_{i\sigma}^{\dagger} c_{j\sigma} +h.c.)+ U\sum_{i}n_{i\uparrow}n_{j\downarrow} \\
 &-A\sum_{i}(s_{i}^{z})^2 -\mu\sum_{i}n_{i}.
\end{aligned}
\end{equation}
The hopping is only between nearest-neighbors, and sites $i$ and $j$ denote in our case the location of cobalts.
Our conclusion is that in order to understand the global spin order, the oxygens can return to its simplistic
role as electronic bridges and we can use Hubbard models based on the TMs only.
On the Hamiltonian above, first we performed a Hubbard-Stratonovich transformation followed by the saddle point approximation for the charge auxiliary field to obtain the spin-fermion model shown below (for details see~\cite{mukherjee2014testing}),

\begin{equation}
\begin{aligned}
H_{SF}&= -t\sum_{\langle i,j \rangle,\sigma} (c_{i\sigma}^{\dagger}c_{j\sigma} + h.c.) +\sum_{i} [-2U\mathbf{m}_{i} \cdot \mathbf{s}_{i} \\ + \frac{U}{2}n_{i}\langle n_{i}\rangle
 &-2Am_{i}^{z}s_{i}^{z}] + \sum_{i}[U|\mathbf{m}_{i}|^2 -\frac{U}{4}\langle n_{i}\rangle^2  \\
 &+ A(m_{i}^{z})^2]  -\mu\sum_{i}n_{i}.
\end{aligned}
\end{equation}

For simplicity, translational symmetry, and because there is no reason for a charge density wave to form, in the local density we assumed $\langle n_{i}\rangle=n$, with $n$ the average electron filling per site i.e. 1 for the half-filling case addressed here to avoid the extra complication of hole or electron doping. We studied the above model by performing classical Monte Carlo (MC) simulations to sample the magnetic moment mean-field order parameter $\mathbf{m}_{i}$, re-diagonalizing the fermionic sector at each MC step. The chemical potential $\mu$ is tuned to the targeted density of electrons. $s_{i}^{z}$ is the $z$-axis spin for the itinerant
fermions, quadratic in the fermionic operators.

We studied $12\times12$ lattices using MC, performing an annealing procedure from high temperature $1.0t$ to low temperature $0.001t$. We fixed the anisotropy $A=0.5t$ and repeated for various values of $U$. At $T=0.001t$, we clearly found zigzag-phase tendencies for $U \gtrapprox 3$, as shown in Fig.~\ref{comparison}. In practice during the annealing process often inhomogenous states were found because of the degeneracy three of the zigzags, with patches of each of the three states separated by domain walls.  For further investigation and to lower the energy, we fixed the orientation of the auxiliary spins as in a perfect zigzag phase and also repeated the procedure for other states that we found are close in energy, such as the $(\pi,0)$ phase (see below) and 120$^{\circ}$-degree phase (either in XZ plane or XY plane) and performed MC on a single parameter $|\mathbf{m}|=|\mathbf{m}_{i}|$ at low temperature $T=0.001t$. We found that for $A=0.5t$ and $1.0t$, the zigzag phase is always the lowest in energy for $U\gtrapprox 3t$. Zigzag states were also reported at intermediate values of $U$ in Ref.~\cite{chern2018semiclassical} using a similar technique.

\begin{figure}
\centering
\includegraphics[width=0.48\textwidth]{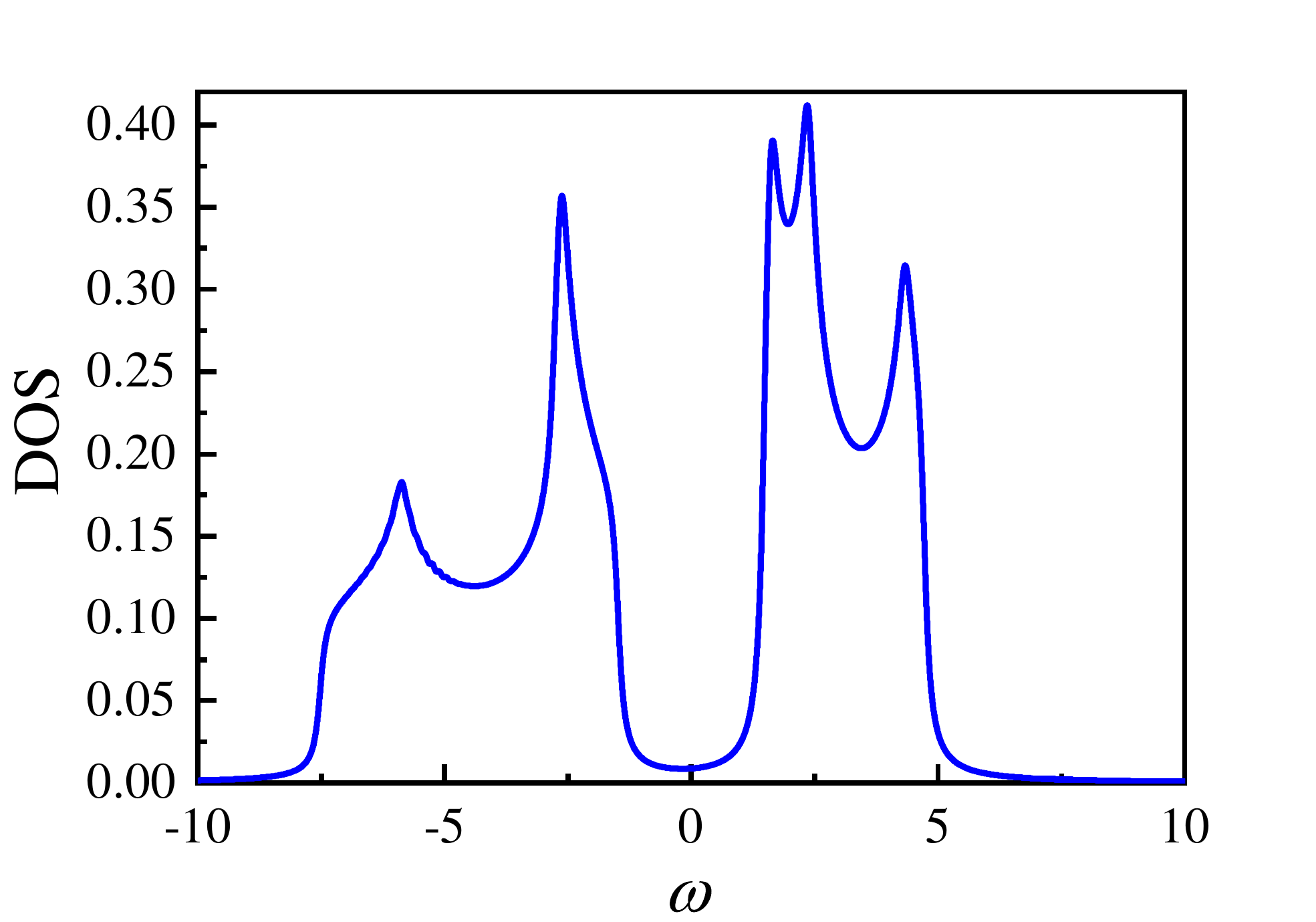}
\caption{Density of states of the zigzag ground state for a $100\times100$ cluster, using the momentum space approach, at $U=6t$ and $A=1.0t$.}
\label{dos_zigzag}
\end{figure}

\begin{figure*}
\centering
\includegraphics[width=0.96\textwidth]{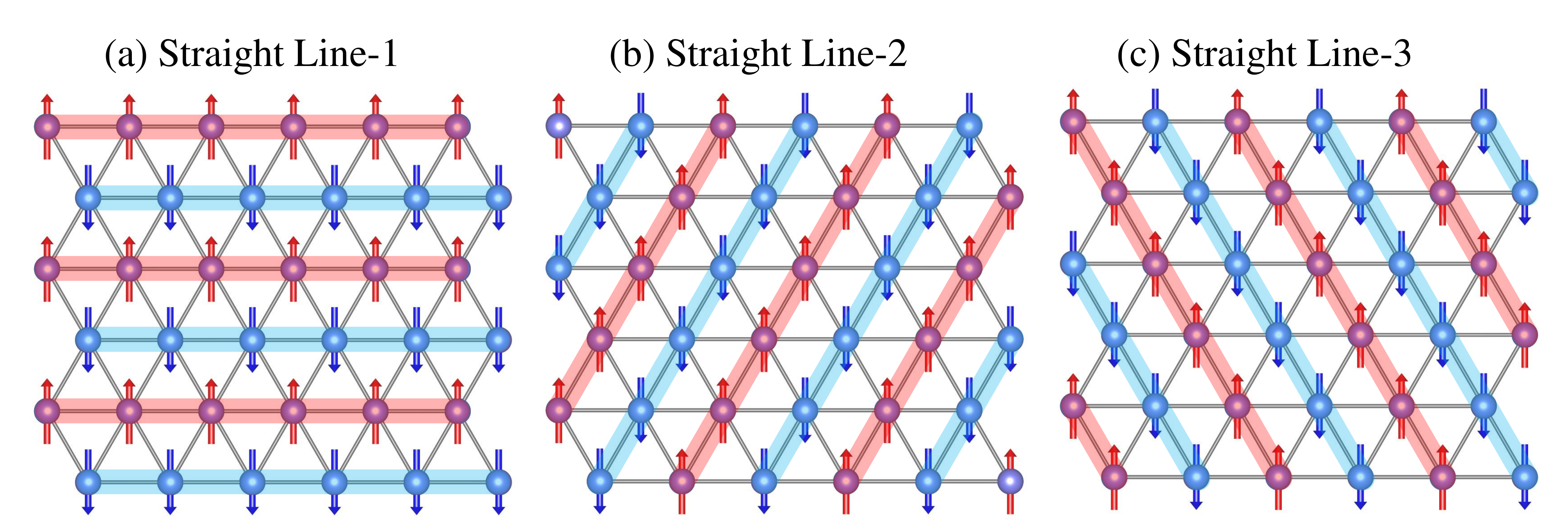}
\caption{Three degenerate $(\pi,0)$ ``straight lines'' states shown in a $6\times6$ cluster. These states are close in energy to the true zigzag ground states, and they
may become ground states under appropriate crystal structure distortions.}
\label{mag_ap}
\end{figure*}

We also performed calculations, at $T=0.001t$, using the Hamiltonian in momentum space. We minimized the energy by tuning $|\mathbf{m}|$ while fixing the magnetic order to the zigzags patterns (see Fig.~\ref{mag_hu}) as well as to the competing $(\pi,0)$ phases (see Fig.~\ref{mag_ap}). This confirmed the MC predictions that zigzags have lower energy that $(\pi,0)$ states. However, it must be noted that as $U/t$ increases the energy gap between the zigzag and $(\pi,0)$ states decreases to very small values, suggesting a high degeneracy at very large $U/t$. Thus, the {\it intermediate} $U/t$ regime is physically the most likely range where the zigzag chains become the ground state. In this region, we should not use a Heisenberg model description to address magnetism in BCO because it will require using exchange couplings $J$ that are FM. Having FM $J$'s can only be phenomenological, and they do not conceptually address the far more complex physics unveiled here.

We also show the single-particle density of states calculated using our momentum-space approach (see Fig.~\ref{dos_zigzag}), for zigzag phases at $U=6.0t$ and employing a $100\times100$ system. This clearly displays a robust gap at the chemical potential establishing that the zigzag phase we found is an insulator, as in experiments. Related zigzag states were also found to be insulators~\cite{chern2018semiclassical}.

For completeness, we close this subsection discussing that the zigzag spin states are also ground states of some multiorbital models unrelated to BCO, as shown in Appendix \ref{subsec:zigzag}. These combinations of results lead us to argue that the introduction of strong easy-axis anisotropy tends to favor zigzag structures under general circumstances within triangular lattices, as it happens in simpler Ising models also in triangular lattices.

\section{Conclusions}
We have shown that our simple model calculations, supplemented by DFT, provide an intuitive explanation for the exotic properties of Ba$_2$CoO$_4$. Firstly, we found the presence of nodes in the spin density along the TM-O bond, which is a consequence of singly-occupied antibonding molecular orbitals. Secondly, the net magnetization exhibited on ligands is found to result from hybridization between atoms and mobility of the electrons with spins {\it opposite} to the spin of the closest TM atoms. In addition, when TM ions are ferromagnetically ordered, regardless of the number of intermediate ligands, a net magnetization will be present in all these ligands, although in some cases with very small values.

Interestingly, for the other case of AFM order on the TM's, the magnetic polarization on ligands depends on the number of those ligands. Specifically, when the path between TM's has an even number of ligands, such as two in the super-super-exchange case TM-O-O-TM that characterizes BCO, a net magnetization should also appear on all the ligands, with different signs. However, for an odd number of ligands between TM's, such as one in the canonical oxide superexchange TM-O-TM prevailing in Cu-O-Cu or Co-O-Co, the net O-polarization should cancel out in the central ligand by mere symmetry while it will be nonzero in the other ligands if there are three or more. Thirdly, based on computer simulations, our results show that the presence of a robust easy-axis anisotropy plays an important role in stabilizing the zigzag-pattern spin order on a triangular lattice, as observed experimentally, by destabilizing the more canonical 120$^{\circ}$ AFM order.

It is important to remark that our results for the oxygen polarization extend also to double perovskites~\cite{taylor2015magnetic,kayser2017magnetic,vasala2015a2b}. Moreover, they extend to other anionic ligands as well, including Cl$^{-}$ in RuCl$_3$, a material recently widely discussed in the context
of exotic spin liquid states~\cite{sandilands2015scattering,banerjee2016proximate}.  The Cl ions in RuCl$_3$ should develop a magnetic moment along the double FM bonds Ru-Cl-Ru of this compound. Note that at low temperatures, zigzag patterns are also observed in the two-dimensional honeycomb lattice of RuCl$_3$.

\section{Acknowledgments}
We are particularly thankful to the late W. Plummer for bringing to our attention the exotic physics of BCO
and for encouraging us to address theoretically its properties. We are also thankful to J. Zhang for comments about
our manuscript.
L.-F.L., N.K., A.C., A.M., and E.D. were supported by the U.S. Department of Energy (DOE), Office of Science, Basic Energy Sciences (BES), Materials Science and Engineering Division. C.S., L.-F.L. and N.K. acknowledge the resources provided by the University of Tennessee Advanced Computational Facility (ACF).

\section{Appendix}
\subsection{\label{subsec:zigzag}Zigzag states in two-orbital double-exchange model}
\begin{figure}
\centering
\includegraphics[width=0.48\textwidth]{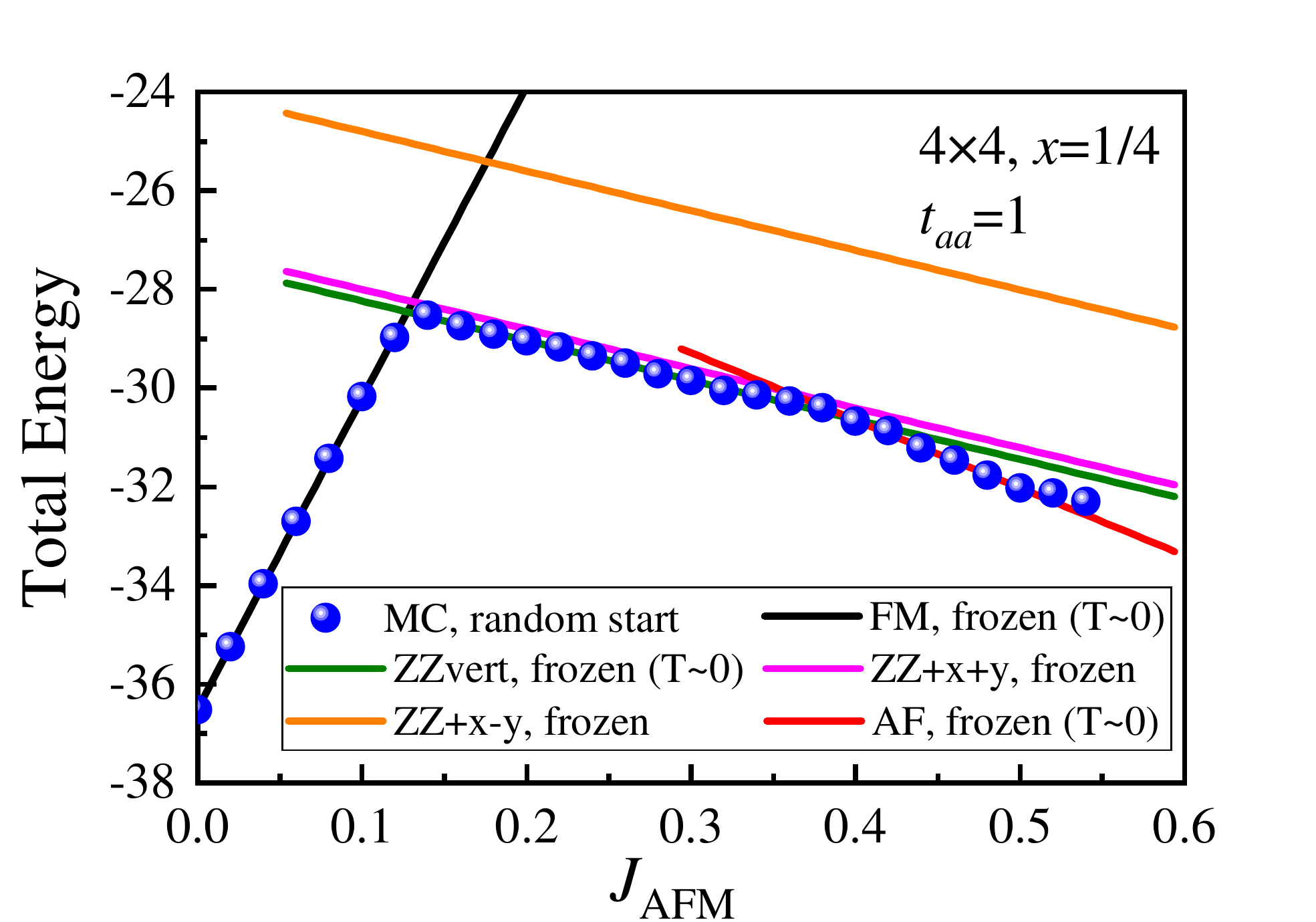}
\caption{Total energy of a $4 \times 4$ cluster double-exchange model as discussed in the text vs. $J_{\rm AFM}$. The blue points are the result of a MC simulation, namely the numerically exact ground state energy for the cluster considered, while the colored lines are the energies of states where the classical $t_{2g}$ spins are frozen to be ferromagnetic (FM), three zigzag (ZZ) states (two of them degenerate), or the 120$^{\circ}$-degree non-collinear AFM state. The two degenerate ZZ states are slightly split for the visual benefit of the readers.}
\label{Figmanga}
\end{figure}

In this Appendix, we will show that zigzag states appear in the context of models for manganites,
illustrating how general our
conclusions are. Recent studies by our group analyzed double-exchange models~\cite{Dagotto:Prp,csen2020properties}, involving the two orbitals $x^2-y^2$ and $3z^2-r^2$ and a classical spin representing the $t_{2g}$ spin. For this reason, i.e. mere simplicity, we chose to study this model to provide additional evidence that our long-range order conclusions are not limited to Ba$_2$CoO$_4$ and are generic for models with strong spin anisotropy.  Also for simplicity, the Jahn-Teller lattice distortions often employed for manganites were not included. The actual Hamiltonian is rather complex and we will not repeat it here explicitly, but instead we refer readers to Eq.(1) of Ref.~\cite{csen2020properties} for its particular
form and for the $2 \times 2$ hopping matrices used.
Note that in Ref.~\cite{csen2020properties}, as well as in here, we included an easy-axis anisotropy $-A \sum_i(S^z_i)^2$, usually not employed in manganites, where the spin is the classical $t_{2g}$ spin.
The inter-orbital hoppings are different by a sign along the $x$- and $y$-axis when  using a  square lattice, with all the other values the same. For the present study, we transformed the square lattice studied in Ref.~\cite{csen2020properties} to the triangular lattice by simply adding an extra hopping along only one of the diagonals of the plaquettes (and we chose the hoppings of that diagonal to be the $x$-axis hoppings, again for mere simplicity).
The resulting model, now on a triangular lattice after adding one plaquette diagonal, was studied by MC simulations. The standard approximation of infinite Hund coupling was used~\cite{Dagotto:Prp}.  The selected electronic density is one electron per site, i.e. half electron per orbital (quarter filling) which is the electronic density of the $e_g$ sector in materials such as LaMnO$_3$. Thus, there is only one parameter to vary, the antiferromagnetic coupling $J_{\rm AFM}$ among the classical spins.

The results of a MC investigation for a $4 \times 4$ cluster are in Fig.~\ref{Figmanga}. We used a low temperature $T=t_{aa}/100$, with $t_{aa}$ the hopping among the $x^2-y^2$ orbitals, employing 10,000 MC steps for thermalization and 50,000 for measurements (keeping only 1 out of 5 lattice configurations to avoid autocorrelation errors). At $J_{\rm AFM}=0$, the double-exchange mechanism is known to favor ferromagnetism in the vicinity of quarter filling, in agreement with the physics of materials such as La$_{1-x}$Sr$_{x}$MnO$_3$~\cite{Dagotto:Prp}. Even undoped LaMnO$_3$ has planar ferromagnetic tendencies, with the effective interlayer coupling being antiferromagnetic (A-type AFM state). Our simulations correctly reproduce this limit. However, as $J_{\rm AFM}$ grows ferromagnetism is penalized, as indicated by the positive slope in the energy of the ferromagnetic state in Fig.~\ref{Figmanga}.

In the other limit of large $J_{\rm AFM}$, the typical 120$^{\circ}$-degree antiferromagnetism of the Heisenberg model on a triangular lattice dominates. Note that if the anisotropy $A$ were made much larger, then the 120$^{\circ}$-degree phase would be suppressed. In practice, the interesting result is that increasing $J_{\rm AFM}$ from the ferromagnetic regime, first we observed a transition to precisely the same zigzag states observed in our mean-field study of the one-orbital section. In the manganite-like model studied here, the degeneracy of this state is only 2, as opposed to the 3 found in the previous subsection, because of the arbitrary selection of hoppings along the diagonal of the plaquettes being those of the $x$-axis. Thus, in each triangle two sides have identical hoppings, while one side has interorbital hoppings of different sign. In the one-orbital model studied before all hoppings in the triangle are the same, leading to degeneracy 3. For this reason, the three zigzag states in the double-exchange model all have the same slope with $J_{\rm AFM}$ but the kinetic energy in such background is different due to the hoppings used.

Regardless of these small details, clearly the two-orbital double-exchange model studied here also has a tendency towards zigzag spin arrangements at {\it intermediate} $J_{\rm AFM}$, providing another example of the generality of our results. Note also that for the zigzags to become the ground state, i.e. having lower energy than states with straight lines of aligned spins, such as those in Fig.~\ref{mag_ap}, the {\it intermediate} coupling range is important. In extreme cases of large $U$, for one orbital, or large $J_{\rm AFM}$, for two orbitals, other states take over or become degenerate with the zigzags. This illustrates the importance of having many-body tools able to analyze with accuracy the challenging intermediate-range coupling which is of much importance to many materials.

\bibliographystyle{apsrev4-1}
\bibliography{ref3}
\end{document}